\documentclass[acmsmall,nonacm]{acmart}
\authorsaddresses{}
\AtBeginDocument{%
  }

\usepackage{amsmath}
\usepackage{algorithm}
\usepackage{algorithmic}

\usepackage{xspace}
\usepackage{graphicx}
\usepackage{subcaption}
\usepackage{float}
\usepackage{graphicx}   % for \scalebox
\usepackage{makecell}   % for \makecell
\usepackage{multirow}   % for \multirow
\usepackage{hyperref}
\usepackage{adjustbox}
\usepackage{booktabs}
\usepackage{amssymb}
\usepackage{enumitem}
\usepackage{fancyhdr}
\usepackage{todonotes}

\usepackage[svgnames]{xcolor}
\usepackage{tikz}
\usetikzlibrary{calc}

\def\Snospace~{\S{}}
\def\Fnospace~{\mbox{Fig.\hspace{0.25em}}}
\def\Tnospace~{\mbox{Tab.\hspace{0.25em}}}
\def\Enospace~{\mbox{Equation\hspace{0.25em}}}

\newcommand{\tinyskip}{\vspace{1pt}}
\newcommand{\mypar}[1]{\tinyskip\tinyskip\noindent\textbf{#1.}\xspace}

\newcommand{\sysname}{DFLOP\xspace}
\newcommand{\maxgain}{3.6\xspace}

\usepackage{xstring}
\usepackage{xcolor}
\usepackage{marginnote}
\usepackage{ifthen} 
\usepackage{soul}
\soulregister\cite7 % check cite version used on overleaf.
\soulregister\ref7      % 
\soulregister\pageref7  % 
\soulregister\eqref7    % 
\soulregister\equation7 %
\definecolor{colorR1}{HTML}{90CAF9} % Blue (맑은 하늘색)
\definecolor{colorR2}{HTML}{EF9A9A} % Red (차분한 붉은색)
\definecolor{colorR3}{HTML}{A5D6A7} % Green (부드러운 초록색)
\definecolor{colorCm}{HTML}{FFF59D} % Yellow (개나리색 - 눈 안 아픈 노랑)
\colorlet{colorR1}{colorR1!40}
\colorlet{colorR2}{colorR2!50}
\colorlet{colorR3}{colorR3!60}
\definecolor{colorDf}{HTML}{333333} % Default
\newcommand{\isrevision}{3}
\newcommand{\currentcolor}{colorDf}
\newcommand{\setrevcolor}[1]{%
  \IfBeginWith{#1}{R1}{\renewcommand{\currentcolor}{colorR1}}{% R1
  \IfBeginWith{#1}{R2}{\renewcommand{\currentcolor}{colorR2}}{% R2
  \IfBeginWith{#1}{R3}{\renewcommand{\currentcolor}{colorR3}}{% R3
  \IfBeginWith{#1}{C}{\renewcommand{\currentcolor}{colorCm}}{% C(Common)
  \renewcommand{\currentcolor}{colorDf}}}}}% 
}
\newcommand{\revisionnote}[1]{%
    \ifthenelse{\equal{\isrevision}{3}}{}{%
        \setrevcolor{#1}%
        \marginnote[\colorbox{\currentcolor}{\color{black}\textbf{#1}}]%
                   {\colorbox{\currentcolor}{\color{black}\textbf{#1}}}%
        [-13pt]%
    }\ignorespaces%
}
\newcommand{\revisionnoteLeft}[1]{
    \ifthenelse{\equal{\isrevision}{3}}{}{%
        \setrevcolor{#1}%
        \reversemarginpar
        \revisionnote{#1}
        \normalmarginpar
    }\ignorespaces%
}

\newcommand{\revisionnoteRight}[1]{%
    \ifthenelse{\equal{\isrevision}{3}}{}{%
        \setrevcolor{#1}%
        \normalmarginpar % force right
        \revisionnote{#1}%
    }\ignorespaces%
}

\newcommand{\revisionnoteCommon}[1]{%
    \ifthenelse{\equal{\isrevision}{3}}{}{%
        \renewcommand{\currentcolor}{colorCm}% 
        \marginnote[\colorbox{\currentcolor}{\color{black}\textbf{#1}}]%
                   {\colorbox{\currentcolor}{\color{black}\textbf{#1}}}%
        [-13pt]%
    }\ignorespaces%
}
\newcommand{\revisionCommon}[2]{%
    \sethlcolor{colorCm}% 
    \ifthenelse{\equal{\isrevision}{0}}{#1}{%
        \ifthenelse{\equal{\isrevision}{3}}{#2}{%
            \ifthenelse{\equal{\isrevision}{1}}{\st{#1}\hl{#2}}%
                {\hl{#2}}}%
        }%
}
\newcommand{\revisionnoteCommonLeft}[1]{%
    \ifthenelse{\equal{\isrevision}{3}}{}{%
        \reversemarginpar % 
        \revisionnoteCommon{#1}% 
        \normalmarginpar % 
    }\ignorespaces% 
}

\newcommand{\revision}[2]{%
    \sethlcolor{\currentcolor}% 
    \ifthenelse{\equal{\isrevision}{0}}{#1}{%
        \ifthenelse{\equal{\isrevision}{3}}{#2}{%
            \ifthenelse{\equal{\isrevision}{1}}{\st{#1}\hl{#2}}%
                {\hl{#2}}}%
        }%
}

\begin{document}

\title{DFLOP: A Data-driven Framework for Multimodal LLM Training Pipeline Optimization}


\author{Hyeonjun An}
\email{hyeonjun.an@yonsei.ac.kr}
\orcid{0009-0007-9617-5583}
\affiliation{
  \institution{Yonsei University BDAI Lab}
  \city{Seoul}
  \country{Republic of Korea}
}

\author{Sihyun Kim}
\email{sihyun.kim@yonsei.ac.kr}
\orcid{0009-0005-3853-618X}
\affiliation{
  \institution{Yonsei University BDAI Lab}
  \city{Seoul}
  \country{Republic of Korea}
}

\author{Chaerim Lim}
\email{chaerim.lim@yonsei.ac.kr}
\orcid{0009-0007-0178-4766}
\affiliation{
  \institution{Yonsei University BDAI Lab}
  \city{Seoul}
  \country{Republic of Korea}
}

\author{Hyunjoon Kim}
\email{wns41559@yonsei.ac.kr}
\orcid{0009-0007-4903-6572}
\affiliation{
  \institution{Yonsei University BDAI Lab}
  \city{Seoul}
  \country{Republic of Korea}
}

\author{Rathijit Sen}
\email{rathijit.sen@microsoft.com}
\orcid{0000-0003-4736-2837}
\affiliation{
  \institution{Microsoft Gray Systems Lab}
  \city{Redmond}
  \state{Washington}
  \country{United States}
}

\author{Sangmin Jung}
\email{tojsm@sktelecom.com}
\orcid{0009-0008-3642-3634}
\affiliation{
  \institution{SK Telecom}
  \city{Seoul}
  \country{Republic of Korea}
}

\author{Hyeonsoo Lee}
\email{hyeonsoo.lee@sktelecom.com}
\orcid{0009-0001-8983-0160}
\affiliation{
  \institution{SK Telecom}
  \city{Seoul}
  \country{Republic of Korea}
}

\author{Dongwook Kim}
\email{kimdw@sktelecom.com}
\orcid{0009-0006-9307-822X}
\affiliation{
  \institution{SK Telecom}
  \city{Seoul}
  \country{Republic of Korea}
}

\author{Takki Yu}
\email{takki.yu@sktelecom.com}
\orcid{0009-0008-4447-1353}
\affiliation{
  \institution{SK Telecom}
  \city{Seoul}
  \country{Republic of Korea}
}

\author{Jinkyu Jeong}
\email{jinkyu@yonsei.ac.kr}
\orcid{0000-0002-4905-9244}
\affiliation{
  \institution{Yonsei University}
  \city{Seoul}
  \country{Republic of Korea}
}

\author{Youngsok Kim}
\email{youngsok@yonsei.ac.kr}
\orcid{0000-0002-1015-9969}
\affiliation{
  \institution{Yonsei University}
  \city{Seoul}
  \country{Republic of Korea}
}

\author{Kwanghyun Park}
\authornote{Corresponding author.}
\email{kwanghyun.park@yonsei.ac.kr}
\orcid{0000-0003-0757-2725}
\affiliation{
  \institution{Yonsei University BDAI Lab}
  \city{Seoul}
  \country{Republic of Korea}
}

% \renewcommand{\shortauthors}{An et al.} % TODO: Every author's full name or just 1st author's name?
% \renewcommand{\shortauthors}{Hyeonjun An, Sihyun Kim, Chaerim Lim, Hyunjoon Kim, Rathijit Sen, Sangmin Jung, Hyeonsoo Lee, Dongwook Kim, Takki Yu, Jinkyu Jeong, Youngsok Kim, Kwanghyun Park} % TODO: Every author's full name or just 1st author's name?
% \renewcommand{\shortauthors}{H. An, S. Kim, C. Lim, H. Kim, R. Sen, S. Jung, H. Lee, D. Kim, T. Yu, J. Jeong, Y. Kim, K. Park.} % TODO: Every author's full name or just 1st author's name?
\renewcommand{\shortauthors}{Hyeonjun An et al.} % FIXED: only 1sst author's full name

\begin{abstract}
 Multimodal Large Language Models (MLLMs) have achieved remarkable advances by integrating text, image, and audio understanding within a unified architecture. However, existing distributed training frameworks remain fundamentally data-blind:  they parallelize computation without accounting for variations in input data characteristics. This data unawareness leads to severe computation skew across stages and microbatches, where heterogeneous multimodal inputs incur different processing costs. Consequently, GPU resources are unevenly utilized, synchronization delays accumulate, and overall training efficiency degrades. To address this limitation, we present \sysname, a \textbf{\underline{d}}ata-driven \textbf{\underline{f}}ramework for multimodal \textbf{\underline{L}}LM training pipeline \textbf{\underline{op}}timization. \sysname continuously profiles runtime behavior to capture data-induced computation variance and employs predictive scheduling to balance workloads across stages and microbatches. By coupling data characteristics with execution planning, \sysname substantially improves GPU utilization and throughput. Extensive experiments on large-scale multimodal benchmarks show that \sysname achieves up to \maxgain {\small$\times$} faster training compared to state-of-the-art distributed training frameworks.
\end{abstract}

%%%


% \keywords{Do, Not, Use, This, Code, Put, the, Correct, Terms, for,
 % TODO do we need to add more?

\begin{comment}

\received{October 2025}
\received[revised]{January 2026}
\received[accepted]{February 2026}
\end{comment}

\maketitle
\begingroup
\renewcommand\thefootnote{}
\footnotetext{Accepted to SIGMOD 2026; to appear in \textit{Proceedings of the ACM on Management of Data (PACMMOD)}.}
\endgroup
\section{Introduction}
Multimodal Large Language Models (MLLMs) have rapidly gained prominence as transformative advancements in artificial intelligence, expanding the capabilities of traditional LLMs by integrating diverse data modalities such as text, image, audio, and so on \cite{efficientmllmsurvey, mplug-owl, qwen, codi, video-llama, mastering}. Their widespread adoption spans various domains, including healthcare, autonomous systems, and human-computer interaction, where their ability to process and generate multimodal content enables new possibilities \cite{multimodallearning, auto, mllmsurvey, agent, healthcare}.

While initial research primarily focused on tasks involving paired single images and text, state-of-the-art models are now designed to handle a far richer and more complex set of inputs, frequently processing multiple images or even video frames within a single instance \cite{llavaov, qwenvl, internvl, llavaov2}. Consequently, modern MLLM training datasets are characterized by a highly heterogeneous mixture of data types. This growing diversity of input data fundamentally complicates the training process \cite{disttrain}. Despite their potential, training MLLMs remains a challenging task. The process demands immense computational resources, intricate model architectures, and the handling of complex multimodal data, often resulting in inefficient GPU utilization and prolonged training times \cite{optimus}. %These challenges pose significant barriers to fully realizing the potential of this transformative technology.

% V1
\mypar{Distributed Training and Its Limitations for MLLMs}
To meet the immense computational and memory demands of large-scale model training, modern frameworks employ a combination of data~\cite{torch_ddp, zero, fsdp}, tensor~\cite{megatron-lm, distmm}, and pipeline~\cite{gpipe, pipedream} parallelism. These three techniques are often combined as 3D parallelism~\cite{megatron-lm}, the de facto paradigm for training trillion-parameter models in frameworks such as Megatron-LM~\cite{megatron-lm}, which build upon PyTorch~\cite{pytorch} primitives. However, these distributed frameworks remain fundamentally data-agnostic. They parallelize computation under an assumption of homogeneous workloads, the assumption violated by the diverse and dynamic nature of multimodal data ~\cite{disttrain, optimus, distmm}. Two critical challenges arise from this mismatch.

A critical challenge in MLLM training is \textit{load imbalance across heterogeneous stages}. While 3D parallelism systems assume consistent microbatch execution times, MLLM pipelines are inherently heterogeneous, comprising architecturally distinct modality encoders and LLMs. Furthermore, processing diverse visual inputs creates data heterogeneity within microbatches. This violates the homogeneity assumption, leading to severe load imbalances. Concurrently, \textit{performance variability stemming from input-dependent throughput} presents another significant hurdle. A model's achievable throughput is highly sensitive to input tensor shapes. This is particularly problematic for MLLMs, where the modality encoder experiences variable \revisionnote{R2.O1}\revision{}{effective batch sizes} and the LLM processes sequences with highly variable lengths. This variability is further compounded by tensor parallelism (TP), where different degrees yield distinct throughputs for identical input shapes. Existing data-agnostic systems fail to model or adapt to this input-dependent behavior, resulting in suboptimal parallelism strategies and reduced end-to-end performance.

\mypar{Towards Data-driven MLLM Training}
While 3D parallelism enables unprecedented scaling, its data-blind design renders it inefficient for MLLM training pipelines. The resulting computation skew and synchronization delays highlight the need for a data-driven framework that adapts distributed training to the characteristics of multimodal data--an approach realized in our proposed system, \sysname.

\sysname is a data-driven framework for optimizing MLLM training pipelines. It explicitly models the heterogeneity of both multimodal architecture and input data to co-optimize parallelism configuration and runtime scheduling, addressing stage imbalance and input-dependent throughput variability. \sysname comprises three major components: (i) Profiling Engine, (ii) Data-aware 3D Parallelism Optimizer, and (iii) Online Microbatch Scheduler. Together, these components bridge data characteristics and distributed execution to achieve balanced and efficient training (\autoref{sec:overview}).

The Profiling Engine quantitatively characterizes both the model and the data workload. It measures the model's memory consumption and throughput across a range of input shapes to construct predictive performance models, and concurrently analyzes the input shape distribution of the actual training dataset (\autoref{sec:profiling_engine}).

The Data-aware 3D Parallelism Optimizer leverages the information provided by the Profiling Engine to determine the 3D parallelism strategy for both the modality encoder and the LLM independently, with the objective of minimizing the expected makespan under the observed data workload (\autoref{sec:optimizer}).

Finally, the Online Microbatch Scheduler addresses uneven computation across individual data items during training. In each step, it calculates estimated computation times and partitions the global batch into microbatches with the objective of balancing the processing load across all pipeline stages, thereby reducing idle GPU time and improving overall throughput (\autoref{sec:scheduler}).

\mypar{Novelty \& Contributions}
Our work introduces a data-driven approach that co-designs the static parallelism strategy and the dynamic runtime scheduling for MLLM training. Unlike prior data-agnostic approaches that optimize solely for a given model architecture, \sysname is the first system, to the best of our knowledge, to mitigate the dual challenges of pipeline imbalance and input-dependent performance variability by explicitly leveraging data distribution statistics and component-level performance profiles. By treating the workload as a first-class component of the optimization problem, our approach mitigates critical structural inefficiencies in large-scale MLLM training pipelines. The main contributions of this paper are as follows:

% \vspace{-10pt}
\begin{itemize} [left=0pt]
\item \textbf{Formalization of MLLM Parallelism Challenges} 
(\autoref{sec:motivation}): We are the first to formally identify and study the dual challenges of stage load imbalance and input-dependent throughput variability in the context of 3D parallelism for MLLMs, demonstrating that data heterogeneity is a critical performance bottleneck.

\item \textbf{The \sysname Framework}
(\autoref{sec:system}): We design and implement \sysname, a novel framework that resolves the dual challenges of MLLM training. \sysname comprises three interconnected components: a Profiling Engine that gathers essential model and data characteristics, a Data-aware 3D Parallelism Optimizer that utilizes this information to determine independent 3D parallelism strategies for the modality encoder and the LLM, and an Online Microbatch Scheduler that operates asynchronously during training to dynamically schedule microbatches and balance workloads in real-time.

\item \textbf{A Flexible and Open-source Parallelism Framework} 
(\autoref{sec:impl}): We implement a new parallelism framework on top of PyTorch that enables separate and independent 3D parallelism for the modality encoder and the LLM. This includes a novel Inter-model Communicator abstraction to efficiently handle communication between disparate data-parallel groups, which we release as open source\footnote{https://github.com/BDAI-Research/DFLOP}. % TODO Should we Change into GitHub?

\item \textbf{Comprehensive Experimental Evaluation}
(\autoref{sec:evalutaion}): We conduct a comprehensive evaluation using state-of-the-art MLLMs and realistic, heterogeneous datasets. Our experiments demonstrate that \sysname substantially outperforms existing baseline systems, improving end-to-end throughput by up to \maxgain{\small$\times$}, without compromising model accuracy.

\end{itemize}
\section{Background and Motivation}
\label{sec:motivation}
\begin{figure*}[t!]
    \centering
    \includegraphics[width=0.99\linewidth,trim={2mm 1mm 4mm 2mm},clip]{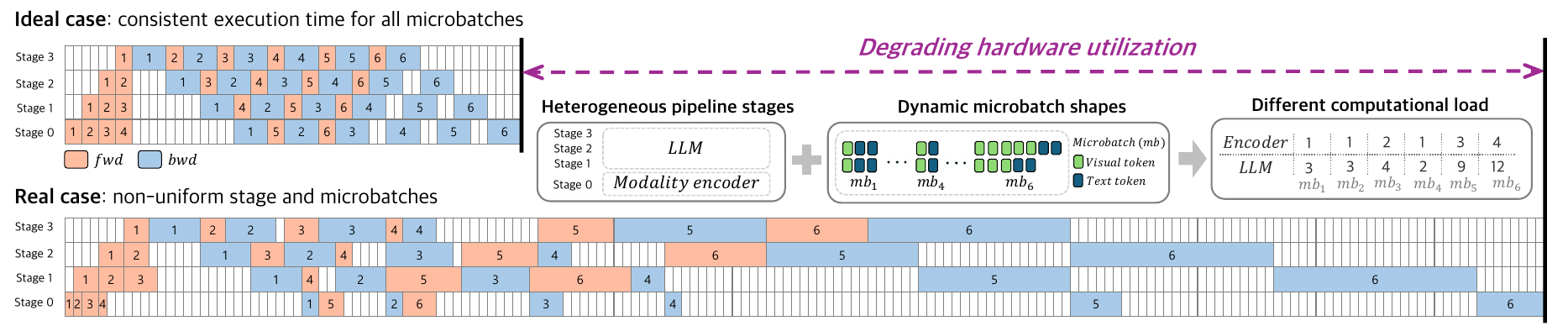}
    % \vspace{-1mm}
    \caption{1F1B~\cite{1f1b} pipeline schedules with backward passes (blue) taking twice as long as the forward pass (pink) with 6 microbatches (represented by numbers) from a mixed dataset consisting of single image~\cite{llava, AI2D, Infographicvqa}, multiple images~\cite{m4-instruct}, and videos~\cite{llava-video} on LLaVA-OV~\cite{llavaov}. The top figure illustrates the ideal 1F1B case, assuming all microbatches have the same execution time. The bottom figure shows the real 1F1B case, where modality encoder is assigned to stage 0 and LLM is assigned to stages 1, 2, and 3. The size of each microbatch in the pipeline varies due to heterogeneous stages and dynamic microbatch shapes. Modality encoder processes only visual tokens, while  LLM processes both visual and text tokens. %Figure 10 언급 (pipeline bubble)}
    }
    %\kp{Typo in Y axis}
    \label{fig:unbalance_pipe}
    \vspace{-5pt}
\end{figure*}

% \begin{figure}[t!]
%     \centering
%     \includegraphics[width=\columnwidth]{figures/mllm_arch.pdf}
%     \caption{Example of an MLLM architecture using a Vision Transformer as a modality encoder.}
%     \vspace{-6mm}
%     \label{fig:mllm_arch}
% \end{figure}

% \begin{figure*}[t]
%     \centering
%     \includegraphics[width=0.99\linewidth]{figures/ddml_motivation.pdf}
%     \label{fig:unbalance_pipe}
% \end{figure*}

% \subsection{MLLM}
% The foundational architecture of contemporary Multimodal Large Language Models (MLLMs) has largely converged on a tripartite structure, consisting of a vision encoder, a projector, and a backbone Large Language Model (LLM). This paradigm initially demonstrated proficiency with single image inputs, but its capabilities have been rapidly extended to more complex data formats. Recent advancements, as highlighted in these works, now enable models to process multiple interleaved images and even continuous video streams, marking a significant expansion in their perceptual scope. Notably, this evolution has been achieved while maintaining the core architectural integrity. Instead of redesigning the fundamental components, innovation has focused on adapting the input processing to support a growing variety of data types and formats. Consequently, this unified architecture is now tasked with interpreting a diverse and heterogeneous stream of visual information, which naturally includes images and video frames of widely varying resolutions.
\begin{figure}[t!]
    \centering
    \begin{minipage}[t]{0.7\linewidth}
        \includegraphics[width=0.99\linewidth,trim={0pt 5pt 0pt 0pt},clip]{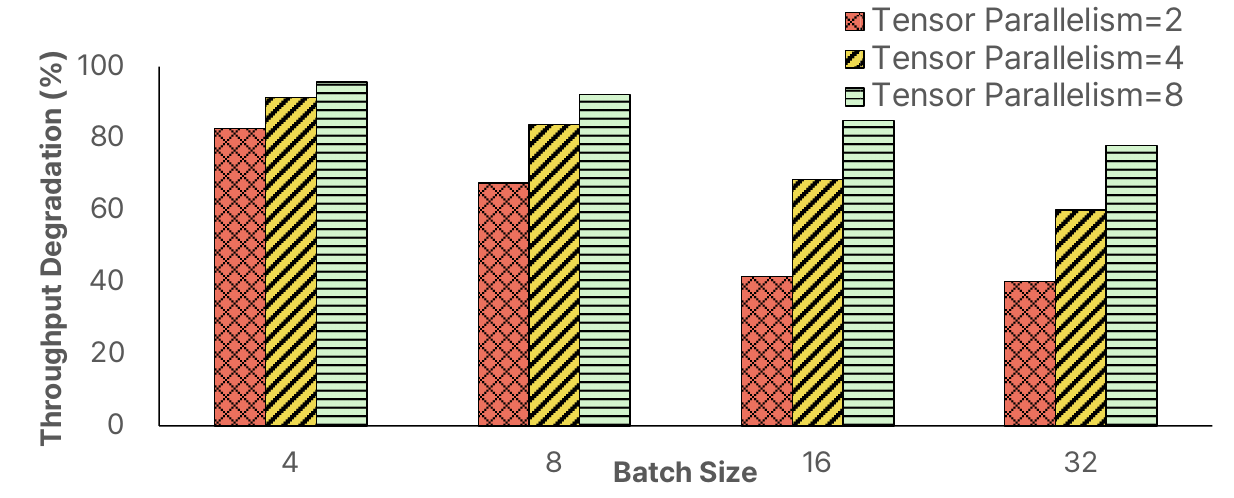}
        \subcaption{\centering Throughput degradation of modality encoder (SigLIP) varying \revisionnote{R2.O1}\revision{}{effective batch size}.}
    \end{minipage}
    \begin{minipage}[t]{0.7\linewidth}
        \includegraphics[width=0.99\linewidth,trim={0pt 5pt 0pt 0pt},clip]{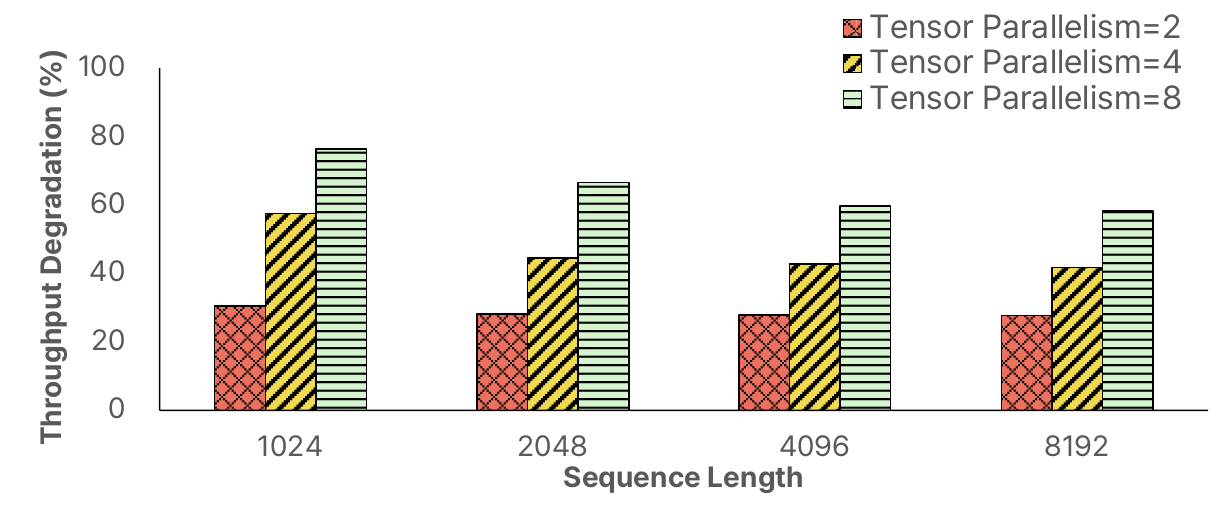}
        \subcaption{\centering Throughput degradation of LLM (Qwen-2.5) varying sequence length.}
    \end{minipage}
    % \vspace{-10pt}
    \caption{Throughput variation with respect to input shape, measured on an 8-GPU HGX A100 node interconnected via NVLink. Throughput degradation with increasing tensor parallelism (TP) arises from computation partitioning. This can lead to GPU underutilization if the resulting workload fragments are too small, compounded by the inherent communication overhead required for synchronization.}
    % \vspace{-12pt}
    \label{fig:input_shape}
% \vspace{-10pt}
\end{figure}

\subsection{MLLM Architecture}
MLLMs are designed to extend the capabilities of traditional LLMs by incorporating multiple data modalities beyond text. MLLMs typically comprise three key submodules: modality encoders, an LLM, and connectors that seamlessly integrate the two components.

Modality encoders are specialized for processing non-textual data such as images, videos, audio, and 3D point clouds by encoding them into vector representations suitable for LLM. Images, sampled video frames, audio data, and 3D point cloud data are preprocessed and encoded to extract visual~\cite{nfnet, vit, clip, eva, beit, pointclip}, auditory~\cite{cformer, hubert, beats}, and spatial~\cite{ulip} features, respectively. LLM~\cite{flant5, qwen, chinchilla, LLaMA1, LLaMA, vicuna} serves as the backbone of the MLLM, processing textual inputs and modality features provided by the encoders to generate textual outputs. Since LLMs are inherently designed to process only textual data, bridging the gap between natural language and other modalities is critical. To address this, a learnable connector is introduced to align the modality encoder outputs with a format compatible with the LLM. These connectors can be implemented in various forms, including linear projectors, multi-layer perceptrons (MLPs), and advanced architectures~\cite{qformer, pformer, mqformer}. 
% \vspace{-2pt}

\subsection{Scaling Visual Inputs in Modern MLLMs}
To address the challenge of processing varied visual inputs, a critical innovation has been the development of dynamic resolution \cite{qwenvl, llavaov, internvl}, a departure from earlier models that uniformly resized images, often at the cost of crucial information. This paradigm enables MLLMs to process images at much higher fidelity and in their native aspect ratios. Orthogonal to these improvements in single-image fidelity, the focus of MLLM development is also expanding beyond single-image contexts. A significant trend is the development of models capable of processing interleaved sequences of multiple images and videos within a single training instance. This evolution is driven by the need to perform more complex reasoning and comprehension tasks that require understanding relationships across multiple visual inputs or over time. Recent models  explicitly incorporate multi-image and video data into their training pipelines \cite{llavaov, llavaov2, internvl}, enabling task transfer between modalities and unlocking new capabilities in areas like comparative reasoning and temporal understanding. This move towards handling heterogeneous, multi-visual inputs represents the next step in building more versatile and capable MLLMs, consequently leading to a substantial increase in the number of visual tokens processed per instance.

\subsection{Motivation}
\label{sec:motivation}
% The training paradigm for MLLMs has evolved to incorporate increasingly diverse data modalities, including mixtures of a single image, multiple images, and video frames. This growing workload heterogeneity fundamentally challenges the design principles of existing 3D parallelism frameworks, which are predicated on a flawed assumption of workload homogeneity. This assumption manifests as two critical, interconnected challenges in MLLM training. The first is load imbalance, caused by non-uniform computational loads from architecturally distinct stages and heterogeneous data within microbatches. This issue is then exacerbated by input-dependent throughput variability, as the achievable throughput of these stages is not a static value but a dynamic function of the shape of the input data.

The training paradigm for MLLMs has evolved to incorporate increasingly diverse data modalities, including mixtures of a single image, multiple images, and video frames. This growing workload heterogeneity fundamentally challenges the design principles of existing 3D parallelism frameworks, which are predicated on a flawed assumption of workload homogeneity. This mismatch leads to critical, interconnected challenges in MLLM training, primarily load imbalance caused by non-uniform computational loads from architecturally distinct stages and heterogeneous data within microbatches, coupled with input-dependent throughput variability, where the achievable throughput of these stages is not a static value but a dynamic function of the shape of the input data.

\mypar{Non-uniform Computation Time Across Stages and Microbatches}
A core premise of pipeline parallelism is the uniform execution time of microbatches, which enables a tightly packed schedule with minimal idle time, as depicted in the ideal case in \autoref{fig:unbalance_pipe}. However, this premise is violated in MLLM training due to two sources of heterogeneity. First, the pipeline stages themselves are architecturally distinct, comprising different modules such as a modality encoder and an LLM. Second, the workload within each microbatch is non-uniform, as a single global batch can contain a mix of data types with varying computational demands. As shown in the real case of \autoref{fig:unbalance_pipe}, this dual heterogeneity results in severe load imbalances across pipeline stages. Systems that do not account for this non-uniformity incur substantial GPU idle time, which significantly inflates the end-to-end training time.

\mypar{Input-dependent Throughput Variability}
A model's throughput in real workloads is not a static value but a complex function of its input tensor's shape, particularly \revisionnote{R2.O1}\revision{}{effective batch size} and sequence length. As  illustrated in \autoref{fig:input_shape}, the throughput degradation when increasing the TP degree from one varies significantly depending on these input dimensions. This variability is critical in MLLM training, where the modality encoder processes inputs with fluctuating \revisionnote{R2.O1}\revision{}{effective batch sizes} and the LLM handles sequences of highly variable lengths. Consequently, a \revisionnote{R2.O8}\revision{}{statically tuned} 3D parallelism strategy must navigate an intricate trade-off space. The potential throughput degradation from a higher TP degree must be balanced against the performance and memory implications of the corresponding data parallel (DP) and pipeline parallel (PP) degrees. Crucially, this optimization cannot be performed based on a single, pre-configured input shape. Instead, it must consider the entire distribution of input shapes present in the actual training dataset to minimize the expected makespan. Existing data-agnostic systems fail to model this complex interplay and thus result in parallelism configurations that are misaligned with the true workload, leading  to suboptimal end-to-end throughput.

\begin{table}[t!]
\caption{Notations.}
% \vspace{-8pt}
\footnotesize
    \centering
    \renewcommand{\arraystretch}{1.2}
    \begin{tabular}{c l}
    \hline
    \textbf{Notations} & \textbf{Description} \\ \hline
    $N_{mb}$             & Number of microbatches\\
    $E_{batch\_size}$   & \revisionnote{R2.O1}\revision{}{Effective batch size} of modality encoder \\
    $E_{seq\_len}$  & Number of visual tokens on modality encoder\\
    $L_{seq\_len}$  & Number of text and visual tokens on LLM.\\
    % $\max (L_{seq\_len})$ & Maximum number of token which LLM accepts \\
    
    % $E_{dp}$        & Degree of data parallelism on Modality Encoder \\
    % $E_{pp}$        & Degree of pipeline parallelism degree on Modality Encoder \\
    % $E_{tp}$        & Degree of tensor parallelism on Modality Encoder \\
    $E_{dp}$, $E_{pp}$, $E_{tp}$ & \makecell[l]{Degrees of data, pipeline, and tensor parallelism \\ on the modality encoder} \\

    % $L_{dp}$        & Degree of data parallelism on LLM \\
    % $L_{pp}$        & Degree of pipeline parallelism degree on LLM \\
    % $L_{tp}$        & Degree of tensor parallelism on LLM \\
    $L_{dp}$, $L_{pp}$, $L_{tp}$ & \makecell[l]{Degrees of data, pipeline, and tensor parallelism \\ on LLM} \\

    $E_l$           & Number of layers on modality encoder \\
    $L_l$           & Number of layers on LLM \\

    $N_{gpus}$      & Total Number of GPUs \\
    $N_{gpu\_node}$ & Number of GPUs per node \\
    $M_{gpu}$       & Memory capacity of a single GPU \\
    $E_{mem}$       & GPU Memory used by modality encoder\\
    $L_{mem}$       & GPU Memory used by LLM\\
    $GBS$           & Global Batch Size\\

    \hline
    \end{tabular}
    %\vspace{5pt}
    % \caption{Symbols used in this paper and their descriptions\kp{any better word than single data?}
    % \kp{need more specific word for Total number of data}}
    % \caption{Symbols used in this paper and their descriptions. Single data refers to one unit of data used in the multi-modal LLM, and number of data refers to the count of such units.}
    \label{tab:symbols}
    \vspace{-15pt}
\end{table}
\section{\sysname}
\label{sec:system}

To address the challenges of load imbalance and performance variability in MLLM training, we designed \sysname, a data-driven framework for optimizing MLLM distributed training. This section details the architecture and core components of our system. The key notations used throughout this section are summarized in \autoref{tab:symbols}. We begin with a high-level overview (\autoref{sec:overview}), followed by a detailed description of our three key components. These are the Profiling Engine for gathering essential performance and data characteristics (\autoref{sec:profiling_engine}), the Data-aware 3D Parallelism Optimizer for determining a \revisionnote{R2.O8}\revision{}{statically tuned} parallelism strategy in an offline phase (\autoref{sec:optimizer}), and the Online Microbatch Scheduler for performing dynamic load balancing at runtime (\autoref{sec:scheduler}).

\subsection{Overview}
\label{sec:overview}

\begin{figure}[t!]
    \centering
    \includegraphics[width=0.8\columnwidth,trim={7pt 0pt 0pt 0pt},clip]{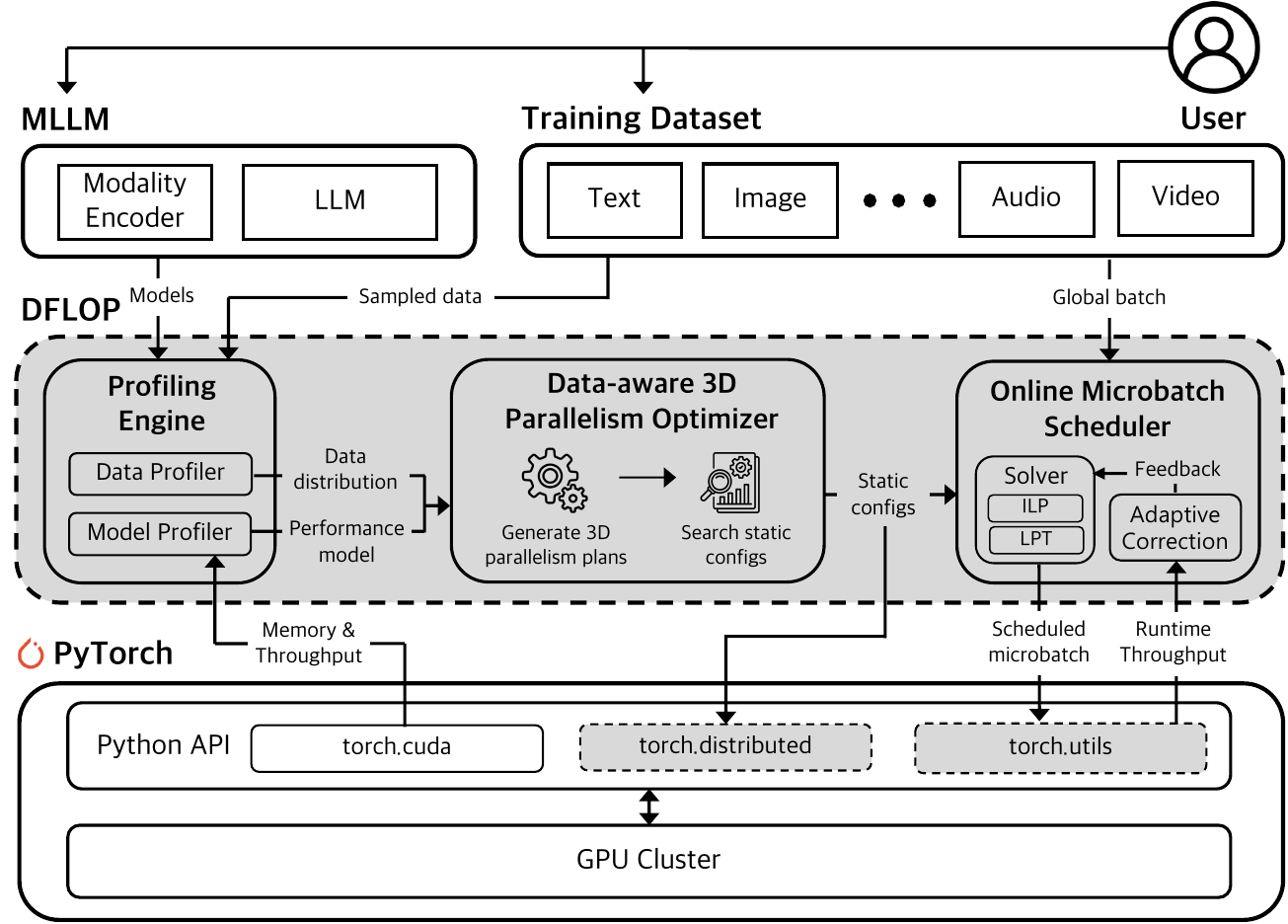}
    % \vspace{-10pt}
\caption{\sysname's three main modules and their integration with an MLLM training framework (PyTorch) shown in gray.}
    \label{fig:sys}
    % \vspace{-15pt}
\end{figure}

This section details the architecture and the operational flow of \sysname, as illustrated in \autoref{fig:sys}. Our system integrates three distinct components to optimize MLLM training. The Profiling Engine and the Data-aware 3D Parallelism Optimizer operate prior to the main training process to gather essential performance characteristics and determine a static parallelization strategy. Subsequently, during the training execution, the Online Microbatch Scheduler operates asynchronously, scheduling the next data batch in parallel with the model's current computation to eliminate scheduling overhead.

The process begins with the Profiling Engine, which takes the user-provided MLLM and training dataset as input. It consists of two sub-components, a Model Profiler and a Data Profiler, that decouple system characterization from workload analysis. The Model Profiler constructs a general, reusable performance model by measuring the throughput and memory consumption of the MLLM on a wide range of synthetic data. Concurrently, the Data Profiler extracts an empirical distribution of input shapes by sampling the specific training dataset.
% This decoupled approach allows the general performance model to be efficiently combined with different workload distributions; optimizing for a new dataset only requires re-executing the faster data profiling step.
The resulting dataset distribution and performance models are then passed to the Data-aware 3D Parallelism Optimizer.

The Data-aware 3D Parallelism Optimizer then uses this information to determine the static configuration. It first generates a comprehensive search space of all possible 3D parallelism strategies for both the modality encoder and the LLM, given the number of available GPUs. It then verifies each configuration against the memory constraints predicted by the performance models. From the configurations that satisfy these constraints, it selects the configuration that minimizes the objective function, which is the expected makespan. This configuration is passed to our custom 3D parallelism framework, built on torch.distributed, which partitions the model and deploys it across the GPU cluster. The configuration details are also forwarded to the Online Microbatch Scheduler.

% The Data-aware 3D Parallelism Optimizer then uses this information to determine the static configuration. It first generates a comprehensive search space of all possible 3D parallelism strategies for both the modality encoder and the LLM, given the number of available GPUs. It then verifies each configuration against the memory constraints predicted by the performance models. From the configurations that satisfy these constraints, it selects the optimal one that minimizes the objective function, which is the expected makespan. This optimal configuration is passed to our custom 3D parallelism framework, built on torch.distributed, which partitions the model and deploys it across the GPU cluster. The configuration details are also forwarded to the Online Microbatch Scheduler.

Finally, during the training execution, the Online Microbatch Scheduler operates. At each iteration, it receives a global batch of data. Using the configuration provided by the optimizer, it predicts the execution duration for each individual data item within the batch. Based on these calculations, it partitions the global batch into the specified number of microbatches in a way that balances the load. This mechanism is integrated into the data loading pipeline, built upon torch.utils.

\subsection{Profiling Engine}
\label{sec:profiling_engine}

The Profiling Engine is an offline component that gathers the empirical data needed for our optimization. It performs two complementary functions. A Model Profiler uses synthetic data to build a general performance model for memory and throughput. In parallel, a Data Profiler analyzes the actual training dataset to create an empirical distribution of its input shapes.

\subsubsection{Model Profiler}
\label{sec:model_prof}
The Model Profiler leverages diverse synthetic data to construct a generalizable performance model spanning a continuous space of input dimensions. This approach characterizes the model's memory and throughput behavior across the entire input spectrum, rather than being coupled to the sparse distribution of a specific training workload.

\revisionnote{R3.O2}\revision{}{We employ empirical profiling as static analytical models fail to capture the non-deterministic execution dynamics inherent to MLLMs. Theoretical formulations often obscure the non-linear performance shifts induced by dynamic input shapes, as illustrated in \mbox{\autoref{fig:input_shape}}, and the scaling disparity between memory-bound attention and compute-bound linear operations~\cite{flashattention, flashattention2}. Furthermore, they remain agnostic to low-level runtime intricacies such as specialized kernels~\cite{pytorch_sdpa, pytorch_tuning}. Capturing these ground-truth characteristics is thus indispensable for ensuring precise load balancing and preventing unexpected OOM errors.}

\mypar{Memory Profiling}  To construct a predictive model for GPU memory consumption, \revisionnote{R3.O2}\revision{}{we first profile the target MLLM across a grid of specific configurations. This grid is defined by varying the number of layers between two distinct small values and scaling the TP degree in powers of two up to $N_{gpu\_node}$. From these measurements, we characterize the total memory footprint as a composition of two primary components: model states and activation states.}

Model states, which encompass parameters, gradients, and optimizer states, are modeled through linear interpolation to derive $model\_state_E(l, E_{tp})$ for the modality encoder and $model\_state_L(l, L_{tp})$ for the LLM, where $l$ denotes the total number of layers. For activation states, which store intermediate results for backpropagation, we account for the distinct input characteristics of each module. While $E_{seq\_len}$ remains fixed for the modality encoder, we employ sequence packing \cite{seq_packing} for the LLM to concatenate instances, effectively fixing the batch size to 1 while making $L_{seq\_len}$ highly variable. Consequently, we model activation memory via linear interpolation based on the \revisionnote{R2.O1}\revision{}{effective batch size} for the modality encoder and sequence length for the LLM. This yields two distinct functions: $act\_state_E(l, E_{tp}, E_{batch\_size}, E_{seq\_len})$ and $act\_state_L(l, L_{tp}, 1, L_{seq\_len})$.
% We decompose the total memory footprint into two primary components: model states and activation states. Model states, which encompass parameters, gradients, and optimizer states, are primarily a function of the number of layers and the TP degree. We build a predictive model for this component by profiling a small reference model across all supported TP degrees and then linearly interpolating based on the number of layers. This process yields two functions: $model\_state_E(l, E_{tp})$ for the modality encoder and $model\_state_L(l, L_{tp})$ for the LLM, where $l$ is the number of layers. Activation states are the intermediate results of the forward pass required for backpropagation. The input shapes for the two modules differ significantly. For the modality encoder, $E_{seq\_len}$ is typically fixed while the $E_{batch\_size}$ varies. To avoid the computational overhead of excessive padding that would result from batching LLM inputs along the batch dimension, we employ sequence packing \cite{seq_packing}. This technique concatenates multiple instances along the sequence dimension, effectively fixing the batch size to 1 while making the $L_{seq\_len}$ highly variable. Consequently, we model activation memory by creating linear interpolation functions based on \rtwo{effective batch size} for the modality encoder and sequence length for the LLM. This yields two distinct functions: one for the modality encoder, $act\_state_E(l, E_{tp}, E_{batch\_size}, E_{seq\_len})$, and another for the LLM, $act\_state_L(l, L_{tp}, 1, L_{seq\_len})$.

\mypar{Throughput Profiling} \revisionnote{R3.O2}\revision{}{To characterize computational efficiency, the Profiling Engine measures the model across a grid of key dimensions: the effective batch size and TP degree for the modality encoder, and the sequence length and TP degree for the LLM.}

For the modality encoder, the function $E_{thr}(E_{batch\_size}, E_{tp})$ is derived by linearly interpolating over the \revisionnote{R2.O1}\revision{}{effective batch size} and TP degree. For the LLM, the core operations including attention and fully-connected layers exhibit distinct scaling behaviors relative to the packed sequence. Attention operations are dependent on individual sequence lengths and must process each original instance separately to maintain causal integrity. In contrast, linear operations depend on the hidden size and can be applied to the entire concatenated sequence at once. To accurately capture these behaviors, these two operations are measured independently and fit using linear interpolation to derive separate throughput models: $L_{attn\_thr}(1, L_{seq\_len}, L_{tp})$ for the attention mechanism and $L_{lin\_thr}(1, L_{seq\_len}, L_{tp})$ for the linear layers.

\subsubsection{Data Profiler}
\label{sec:data_analysis}
The specific dimensions that vary and their impact on the model's inputs are strictly governed by the MLLM's architecture and its data processing pipeline. For instance, some models partition vision data into a variable number of fixed-size patches based on resolution \cite{llavaov, llava-1.5}, while others employ a variable image sequence length \cite{qwenvl}. Similarly, certain architectures preserve the full sequence dimension of modality vectors, whereas others utilize connectors to reduce this dimension to alleviate the computational burden on the LLM \cite{internvl}. These model-specific transformations imply that even with the same dataset, the input shape distributions for the core modules can differ drastically. \revisionnoteLeft{R3.O2}\revision{}{To characterize this workload, the Data Profiler first identifies the varying input dimensions for both the modality encoder and the LLM. It then performs random sampling across the dataset, calculating the precise input shapes for each sampled item within the target architecture to construct empirical histograms that represent the statistical distribution of the workload.}
\subsubsection{Re-profiling Conditions}
\label{sec:reprof_cond}
\revisionnoteCommon{R1.Q1}\revisionCommon{}{The necessity for re-profiling is determined} \revisionnoteCommon{R3.O2}\revisionCommon{}{by changes in the structural or statistical characteristics of the workload. The Model Profiler requires re-profiling only when the model architecture is modified, as its performance model is built upon the hardware-specific execution behavior of fixed operations. In contrast, the Data Profiler must be re-executed when either the model architecture or the underlying dataset changes, as the input shape distributions are jointly determined by the model's preprocessing logic and the raw data characteristics.}
% For a fixed model and dataset, both profiling processes are performed only once as an offline step.

\subsection{Data-aware 3D Parallelism Optimizer}
\label{sec:optimizer}
% \rtwo{The Data-aware 3D Parallelism Optimizer serves as an offline component that establishes the static configuration for the MLLM. Its primary function is to identify a parallelism strategy, including $E_{tp}, E_{pp}, E_{dp}$ for the encoder and $L_{tp}, L_{pp}, L_{dp}$ for the LLM, along with the number of microbatches per pipeline stage $N_{mb}$. The optimizer selects the configuration that minimizes the expected makespan by leveraging the input shape statistics of the training dataset. This selection process is constrained by the total number of available GPUs and the memory capacity of each individual GPU to ensure hardware feasibility. By utilizing the throughput and memory performance models provided by the Profiling Engine, the optimizer effectively explores the search space to determine the most efficient parallel execution strategy for the given hardware and data characteristics.}
\revisionnote{R2.O3}\revision{}{The Data-aware 3D Parallelism Optimizer identifies a static parallelism}\revisionnote{R2.O4}\revision{}{ strategy, comprising $E_{tp}, E_{pp}, E_{dp}$ for the encoder, $L_{tp}, L_{pp}, L_{dp}$ for the LLM, and the number of microbatches per} \revisionnote{R2.O5}\revision{}{  pipeline stage $N_{mb}$. It minimizes the expected makespan subject to the constraints on per-GPU memory capacity and the total number of GPUs, leveraging the throughput and memory models from the Profiling Engine and the training dataset's input statistics provided by the Data Profiler.}

\subsubsection{Problem Formulation}
The primary objective of the optimizer is to find a 3D parallelism configuration that minimizes the training iteration time, or makespan. The makespan ($T$) of a pipelined execution is determined by the $N_{mb}$, the pipeline depth, and the duration of the slowest stage. For an MLLM, the total pipeline depth is the sum of the respective pipeline degrees ($E_{pp}+L_{pp}$), and the makespan is defined as:
% {
% % \small
% $$T=(N_{mb} + E_{pp} + L_{pp} -1) \cdot \max \left(E_{dur}, L_{dur} \right)$$
% }

{
\begin{align*}
    T=(N_{mb} + E_{pp} + L_{pp} -1) \cdot \max \left(E_{dur}, L_{dur} \right)
\end{align*}
}

Here, $E_{dur}$ and $L_{dur}$ represent the execution durations for a microbatch on the modality encoder and the LLM. We define a parameter vector $\theta=(E_{tp},E_{pp},E_{dp},L_{tp},L_{pp},L_{dp},N_{mb})$ to represent a complete 3D parallelism strategy. 
For data $d$ in the sampled dataset $D$, the duration for each module is then expressed as:

{
\begin{align*}
    E_{dur}(d;\theta) = \frac{E_{FLOP}(d;\theta)}{E_{thr}(b(d), E_{tp})}, ~~ L_{dur}(d;\theta)=\frac{L_{FLOP}(d;\theta)}{L_{thr}(s(d), L_{tp})}
\end{align*}
}

% $$E_{dur}(d;\theta) = \frac{E_{FLOP}(d;\theta)}{E_{thr}(b(d), E_{tp})}, L_{dur}(d;\theta)=\frac{L_{FLOP}(d;\theta)}{L_{thr}(s(d), L_{tp})}$$
% {
% \small
% $$E_{dur}(d; E_{tp},E_{pp},E_{dp}) = \frac{E_{FLOP}(d;E_{tp},E_{pp},E_{dp})}{E_{thr}(b(d), E_{tp})}, L_{dur}(d;L_{tp},L_{pp},L_{dp})=\frac{L_{FLOP}(d;L_{tp},L_{pp},L_{dp}}{L_{thr}(s(d), L_{tp})}$$
% }

The throughput functions, $E_{thr}$ and $L_{thr}$, depend on the input shape, specifically the modality \revisionnote{R2.O1}\revision{}{encoder's effective batch size} $b(d)$ and the LLM's sequence length $s(d)$, as well as the chosen tensor parallelism degrees. Because both the computational load per GPU ($E_{FLOP}(d;\theta)$, $L_{FLOP}(d;\theta)$) and the throughput functions ($E_{thr},L_{thr}$) are dependent on each data item $d$, the stage durations ($E_{dur},L_{dur}$) are not static values. Consequently, the makespan, which we denote as $T(d;\theta)$, is also a variable dependent on the specific data item being processed. A data-agnostic approach that relies on a single point-estimate for the workload is therefore insufficient. Instead, our objective is to find the parameter vector $\theta^*$ that minimizes the expected makespan over the entire distribution of the sampled training dataset $D$. This is formulated as:

{
\begin{equation}
    \label{eq:objective}
    \theta^* = \arg\min_{\theta\in\Theta} \frac{1}{|D|} \sum_{d\in D} T(d; \theta)
\end{equation}
}

The optimization is performed over a set of valid configurations $\Theta$, defined by hardware and memory constraints. The TP degrees are typically limited to GPUs within the same node, as tensor parallelism requires high-bandwidth, low-latency communication.

\begin{equation}
    \label{eq:tp}
    E_{tp}\in \{1, 2, \ldots, N_{gpu\_node}\}, ~L_{tp}\in \{1, 2, \ldots, N_{gpu\_node}\}
\end{equation}

The total number of GPUs utilized must equal the number available.

\begin{equation}
    \label{eq:gpu}
    E_{pp}\cdot E_{tp}\cdot E_{dp}+L_{pp}\cdot L_{tp}\cdot L_{dp} = N_{gpus}
\end{equation}

Finally, the memory footprint for each module must not exceed the per-GPU memory capacity, $M_{gpu}$. The memory model incorporates both model states and activation states. A critical consideration is that the activations for the modality encoder must be retained in memory for the duration of the entire pipeline, making their memory cost proportional to the total pipeline depth, $E_{pp}+L_{pp}$. The memory constraints for the modality encoder and the LLM are therefore:
% Add negative vspace to reduce the gap between algorithm 1 and the equation below: 
% \vspace{-20pt}

{\small
\begin{flalign}
    \label{eq:v_mem}
    Mem_E(p_{E}) &= model\_state_E\left(\frac{E_l}{E_{pp}}, E_{tp}\right)
    + (E_{pp}+L_{pp}) \cdot act\_state_E\left(\frac{E_l}{E_{pp}}, E_{tp}, E_{batch\_size}, E_{seq\_len}\right) 
    \leq M_{gpu}, && 
\end{flalign}
where $p_{E} = \{E_{tp}, E_{pp}, E_{batch\_size}, E_{seq\_len}, L_{pp}\}$.
}

{\small
\begin{flalign}
    \label{eq:l_mem}
    Mem_L(p_{L}) &= model\_state_L\left(\frac{L_l}{L_{pp}}, L_{tp}\right)
    + L_{pp} \cdot act\_state_L\left(\frac{L_l}{L_{pp}}, L_{tp}, 1, L_{seq\_len}\right) \leq M_{gpu} 
\end{flalign} 
where $p_{L} = \{L_{tp}, L_{pp}, L_{seq\_len}\}$. \\
}

\noindent Consequently, the Data-aware 3D Parallelism Optimizer aims to find $\theta^*$ that minimizes the expected makespan in Equation \ref{eq:objective}, subject to the hardware and memory constraints defined by Equations \ref{eq:tp}, \ref{eq:gpu}, \ref{eq:v_mem}, and \ref{eq:l_mem}.

\begin{algorithm}[t]
\caption{Find Data-aware MLLM 3D Parallelism Configuration}
\label{alg:opt}
\small
\begin{algorithmic}[1]
    \REQUIRE $N_{gpus}$, $M_{gpu}$, $Profiled.Data$, $GBS$
    \ENSURE $\theta^* = (E_{tp}^*, E_{pp}^*, E_{dp}^*, L_{tp}^*, L_{pp}^*, L_{dp}^*, N_{mb}^*)$
    \STATE \text{\# Phase 1: Generate all possible 3D parallelism configurations}
    % \STATE \text{//} $P_{configs} = \{ (V_{tp}, V_{pp}, V_{dp}, L_{tp}, L_{pp}, L_{dp}) \}$
    \STATE $P_{configs} \gets []$ % needs  \usepackage{amssymb} to use \varnothing % change config to cfg
    % \FOR{$V_{gpus} = 1$ to $N_{GPUs}-1$}
    \FOR{$E_{gpus} \gets 1$ \textbf{to} $N_{gpus} - 1$}
        \STATE $L_{gpus} \gets N_{gpus} - E_{gpus}$

        %%%% prev
        % \STATE $E_{configs} \gets \texttt{FindCombs}(V_{gpus})$
        \STATE $E_{configs} \gets FindCombs(E_{gpus})$ 
        \STATE \text{\footnotesize \# Find all 3D parallel combinations (TP, PP, DP) whose product is E\_gpus}
        % \STATE\text{ // Find all} $\left(V_{tp}, V_{pp}, V_{dp}\right)\in\mathbb{N}^3$ \text{s.t.} $V_{tp}\cdot V_{pp} \cdot V_{dp} = V_{gpus}$

        % Need to import \usepackage{amsfonts}
        \STATE $L_{configs} \gets FindCombs(L_{gpus})$
        \STATE \text{\footnotesize \# Find all 3D parallel combinations (TP, PP, DP) whose product is L\_gpus}
        %%% prev

        %%%% fixed
        % \STATE $(E_{cfgs}, L_{cfgs}) \gets (FindCombs(E_{gpus}), FindCombs(L_{gpus}))$ 
        % \STATE \text{\footnotesize \# Find all 3D parallel combinations whose product is E\_gpus or L\_gpus}
        % % \STATE\text{ // Find all} $\left(V_{tp}, V_{pp}, V_{dp}\right)\in\mathbb{N}^3$ \text{s.t.} $V_{tp}\cdot V_{pp} \cdot V_{dp} = V_{gpus}$
        %%%% fixed

        %%% prev
        \STATE $P_{configs}.add(E_{configs} + L_{configs})$
        %%%
        
        %%%fix
        % \STATE $P_{cfgs}.add(E_{cfgs} + L_{cfgs})$
        %%%
    \ENDFOR
    \STATE

    \STATE \text{\# Phase 2: Find the configuration $\theta^*$ which minimizes T}
    % \STATE $min\_T \gets + \infty$
    % \STATE $\theta^* \gets \text{NULL}$
    \STATE $min\_T \gets +\infty, ~~ \theta^* \gets  \text{NULL}$
    % \STATE $(E_{thr}^*, L_{thr}^*) \gets (0, 0)$
    % \STATE $(min\_T, \theta^*, E_{thr}^*, L_{thr}^*) \gets (+\infty, \text{NULL}, 0, 0)$
    % \STATE $\left(max\_bsz, max\_seq\_len \right) \gets Profiled.Data.max()$
    % add mean value of batch size and seq len
    % \STATE $( bsz_{\mathrm{mean}}, seq\_len_\mathrm{mean}) \gets Profiled.Data.mean()$ 
    \STATE $\left(mean\_bsz, mean\_seq\_len\right) \gets Profiled.Data.mean()$ 

    % \STATE $t\_batch\_size, t\_seq\_len \gets \text{SetMaxInputThresholds}(ratio)$
    % \FOR{$P_{conf} \in P_{configs}$}
    \FOR{$(E_{tp}, E_{pp}, E_{dp}, L_{tp}, L_{pp}, L_{dp}) \textbf{ in } P_{configs}$}
        % \STATE $(V_{tp}, V_{pp}, V_{dp}, L_{tp}, L_{pp}, L_{dp}) \gets P_{conf}$
        % \STATE $V_{mem} \gets \text{CalMem}_V(V_{tp}, V_{pp}, t\_batch\_size)$

        % \textit{// CalMem calculates memory usage based on equation}
        
        % \STATE $L_{mem} \gets \text{CalMem}_L(L_{tp}, L_{pp}, t\_seq\_len)$
        % \IF{$V_{mem} > \text{GPU\_Memory}$ \textbf{or} $L_{mem} > \text{GPU\_Memory}$}
        %     \STATE \textbf{continue}
        % \ENDIF
        \STATE $N_{max\_mbatch} \gets GBS // L_{dp}$
        \FOR{$i = 1$ to $N_{max\_mbatch}$}
            % add target_batch_size and target_seq_len 
            % \STATE $E_{mbatch\_size} = GBS / (i \cdot E_{dp})$
            % \STATE $L_{mbatch\_size} = GBS / (i \cdot L_{dp})$
            % \STATE $t\_bsz = mean\_bsz \cdot E_{mbatch\_size}$
            % \STATE $t\_seq\_len = mean\_seq\_len \cdot L_{mbatch\_size}$
            \STATE $t\_bsz = mean\_bsz \cdot GBS / (i \cdot E_{dp})$
            \STATE $t\_seq\_len = mean\_seq\_len \cdot GBS / (i \cdot L_{dp})$
            \STATE $E\_mem \gets Mem_E\left(E_{tp}, E_{pp}, t\_bsz, E_{seq\_len}, L_{pp}\right)$
            \STATE $L\_mem \gets Mem_L\left(L_{tp}, L_{pp}, t\_seq\_len\right)$
            \IF{$E\_mem > M_{gpu}$ \OR $L\_mem > M_{gpu}$}
                \STATE $\textbf{continue}$
            \ENDIF
            % add end
            \STATE $E_{dur} \gets E_{FLOP} / \left(E_{thr}(t\_bsz, E_{tp})\cdot E_{tp} \cdot E_{pp} \right)$
            \STATE $L_{dur} \gets L_{FLOP} / \left(L_{thr}(t\_seq\_len, L_{tp})\cdot L_{tp} \cdot L_{pp} \right)$
            % \STATE $E_{dur} \gets E_{FLOP} / \left(E_{thr}(t\_bsz, E_{tp})\cdot E_{tp} \cdot E_{pp} \cdot E_{dp}\right)$
            % \STATE $L_{dur} \gets L_{FLOP} / \left(L_{thr}(t\_seq\_len, L_{tp})\cdot L_{tp} \cdot L_{pp} \cdot L_{dp}\right)$
            % \STATE $slowest\_stage \gets \max\left(E_{dur}, L_{dur}\right)$
            % \STATE $T \gets (i + E_{pp} + L_{pp} - 1) \cdot slowest\_stage$
            \STATE $T \gets (i + E_{pp} + L_{pp} - 1) \cdot \max\left(E_{dur}, L_{dur}\right)$
            \IF{$T < min\_T$}
                \STATE $min\_T \gets T$
                \STATE $\theta^* \gets (E_{tp}, E_{pp}, E_{dp}, L_{tp}, L_{pp}, L_{dp}, i)$
                % \STATE $(E_{thr}^*, L_{thr}^*) \gets (E_{thr}(t\_bsz, E_{tp}), L_{thr}(t\_seq\_len, L_{tp}))$
            \ENDIF
        \ENDFOR
    \ENDFOR
    \STATE \textbf{return} $\theta^*$

\end{algorithmic}
\end{algorithm}
% \vspace{-20pt}
%%% SK's fixed algorithm end
\subsubsection{Parallelism Optimization}
\label{sec:parallel_opt}
To solve the optimization problem, we employ the search algorithm detailed in \autoref{alg:opt}. The algorithm proceeds in two main phases. The first phase enumerates the entire search space. It begins by iterating through all possible ways to partition the available GPUs ($N_{gpus}$) between the modality encoder and the LLM. For each partition, the algorithm then identifies all valid 3D parallelism combinations including TP, PP, and DP for each module to create a comprehensive set of candidate strategies $P_{configs}$.

The second phase evaluates this space to select the configuration that minimizes the expected makespan. The algorithm iterates through each candidate in $P_{configs}$ and calculates the theoretical maximum number of microbatches $N_{max\_mbatch}$. It then initiates an inner loop that iterates through each possible microbatch count $i$. Within this loop, the algorithm performs a memory feasibility check. Using the performance models from the Profiling Engine, it estimates the peak memory consumption for both modules denoted as $E_{mem}$ and $L_{mem}$. Any configuration predicted to exceed the GPU memory capacity $M_{gpu}$ is discarded. For memory-feasible instances, the algorithm calculates the makespan $T$ using the average FLOP from the dataset analysis in conjunction with the profiled throughput models to approximate expected stage durations. The algorithm records the configuration $\theta^*$ that yields the minimum estimated makespan across all evaluated candidates and returns this as the selected strategy.

\mypar{Parallelism Optimization Time Complexity Analysis} 
\revisionnoteCommon{R2.O2}\revisionCommon{}{The time complexity of Algorithm 1 stems from search space generation}\revisionnoteCommon{R3.O3}\revisionCommon{}{ and configuration evaluation. The generation phase depends on the number of valid configurations, bounded by the divisor function $d(n) \le O(n^\epsilon)$ for any arbitrarily small $\epsilon > 0$ ~\cite{Tao}. Summing the products of divisors across all GPU partitions yields a total search space size of $O((N_{gpus})^{1+\epsilon})$. Since the evaluation phase iterates through microbatch counts up to the $GBS$, the overall complexity is $O(GBS \cdot (N_{gpus})^{1+\epsilon})$. This sub-polynomial growth ensures that the optimization overhead remains negligible even for large clusters.}

\begin{figure}[t]
  \centering
  \begin{subfigure}[t]{0.4\linewidth}
    \centering
    \includegraphics[width=\linewidth,trim={18pt 16pt 13pt 7pt},clip]{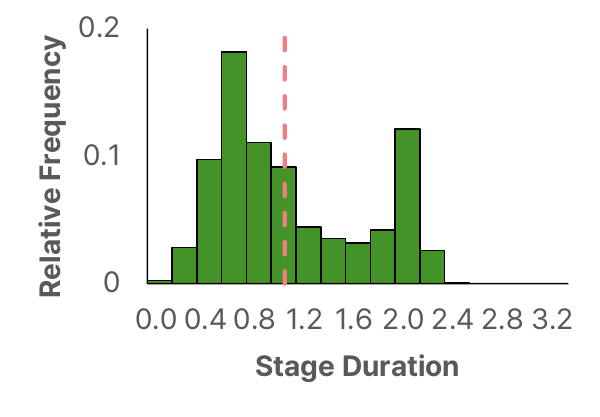}
    \subcaption{Stage duration of modality encoder.}
    \label{fig:stage_duration_vision}
  \end{subfigure}
  \begin{subfigure}[t]{0.4\linewidth}
    \centering
    \includegraphics[width=\linewidth,trim={10pt 16pt 13pt 7pt},clip]{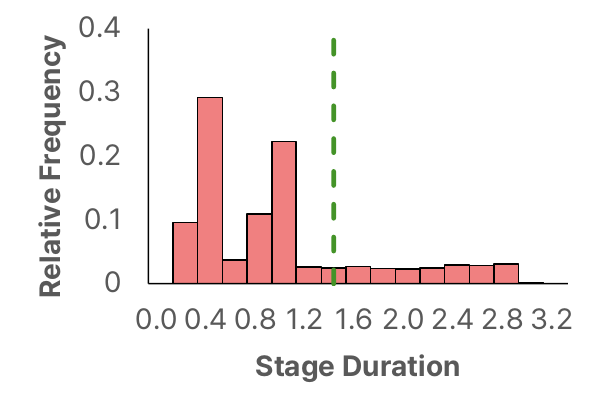}
    \subcaption{Stage duration of LLM.}
    \label{fig:stage_duration_llm}
  \end{subfigure}
  % \vspace{-10pt}
  \caption{Stage-wise duration distributions across data items for (a) modality encoder (SigLIP)  and (b) LLM (Qwen-2.5) stages under a mixed dataset. Vertical lines indicate the mean duration.}
  \label{fig:stage_duration}
  \vspace{-10pt}
\end{figure}
\subsection{Online Microbatch Scheduler}
\label{sec:scheduler}

The Data-aware 3D Parallelism Optimizer establishes a static configuration based on dataset statistics. \revisionnote{R2.O3}\revision{}{However, this approach cannot } \revisionnote{R2.O4}\revision{}{account for the workload variations that occur at runtime within individual batches. During training, the pipeline divides the global } \revisionnote{R2.O5}\revision{}{batch into $m$ buckets where $m$ is defined as the product of $N_{mb}$ and $L_{dp}$. Existing scheduling strategies assign data items to these buckets in a random manner. As demonstrated in \mbox{\autoref{fig:stage_duration}}, this random assignment results in high variability in stage durations due to the heterogeneity of input data shapes. Consequently, this variability induces pipeline bubbles.}

To address this dynamic workload variance, we introduce the Online Microbatch Scheduler. It dynamically partitions the global batch items into $m$ buckets to minimize the variance in stage durations. By leveraging per-instance duration calculations, the scheduler actively balances the computational load across all pipeline stages to minimize pipeline bubbles and maximize hardware utilization.

% While the Data-aware 3D Parallelism Optimizer establishes a statically optimal configuration based on global dataset statistics, it cannot account for the workload variations that occur at runtime within individual batches. The data distribution in any given batch can deviate significantly from the dataset's average, leading to transient load imbalances between pipeline stages, as demonstrated in \autoref{fig:stage_duration}. Such imbalances create pipeline bubbles, which are idle periods where GPUs wait for the slowest microbatch to complete, thereby degrading computational efficiency.

% To mitigate this dynamic, fine-grained workload variance, we introduce the Online Microbatch Scheduler. This component operates as a centralized scheduler at runtime, complementing the offline optimizer by dynamically scheduling the data within each global batch before it enters the training pipeline. Its objective is to partition the incoming data items into microbatches in a manner that balances the computational load across all pipeline stages. By leveraging per-instance performance predictions, the scheduler actively minimizes pipeline bubbles and maximizes hardware utilization for the specific data being processed in each iteration.

\subsubsection{Problem Formulation}
The core task of the Online Microbatch Scheduler is to solve a load-balancing problem for each incoming global batch. For a given global batch consisting of $N$ data items, the goal is to partition these items into $m$ distinct buckets to minimize the pipeline's makespan for that iteration. We formulate this partitioning task as an Integer Linear Programming (ILP) problem. The objective is to minimize the bottleneck, represented by $C_{max}$, which is the maximum execution time across any stage of any microbatch. The complete formulation is as follows:

%change m's definition. it's N_mb times L_dp%
% { \small
% \setlength{\abovedisplayskip}{4pt}
% \setlength{\belowdisplayskip}{4pt}
% \begin{equation}
%     \begin{aligned}
%     & {\textbf{minimize}}
%     % & & C_{max} = \max_{j \in \{1, \dots, N\}} \{E_j, L_j\} \\
%     % & & C_{max} = \max_{1\leq j \leq N} \max \left\{ E_j, L_j\right\} \\
%     & & C_{max} = \max \left( \max_{1\leq j \leq N} E_j,\max_{1\leq j \leq N} L_j \right) \\
%     & \textbf{subject to:}
%     & & \sum_{j=1}^{N} x_{ij} = 1, & \forall i \in \{1, \dots, N\} \\
%     % & & & \sum_{i=1}^{N} E_{\text{dur}_i} \cdot x_{ij} = E_{j}, & \forall j \in \{1, \dots, N_{mb}^*\} \\
%     & & & \sum_{i=1}^{N} E_{dur}(d_i; \theta^*) \cdot x_{ij} = E_{j}, & \forall j \in \{1, \dots, m\} \\ß
%     % & & & \sum_{i=1}^{N} L_{{dur}_i} \cdot x_{ij} = L_{j}, & \forall j \in \{1, \dots, N_{mb}^*\} \\
%     & & & \sum_{i=1}^{N} L_{dur}(d_i; \theta^*) \cdot x_{ij} = L_{j}, & \forall j \in \{1, \dots, m\} \\
%     & & & x_{ij} \in \{0, 1\}, & \forall i, j
%     \end{aligned}
% \end{equation}
% }

\vspace{-10pt}

{
\small
\begin{align}
    % \begin{aligned}
    & {\textbf{minimize}}
    % & & C_{max} = \max_{j \in \{1, \dots, N\}} \{E_j, L_j\} \\
    % & & C_{max} = \max_{1\leq j \leq N} \max \left\{ E_j, L_j\right\} \\
    & & C_{max} = \max \left( \max_{1\leq j \leq m} E_j,\max_{1\leq j \leq m} L_j \right)  \\
    & \textbf{subject to:}
    & & \sum_{j=1}^{m} x_{ij} = 1, & \forall i \in \{1, \dots, N\}  \nonumber \\
    % & & & \sum_{i=1}^{N} E_{\text{dur}_i} \cdot x_{ij} = E_{j}, & \forall j \in \{1, \dots, N_{mb}^*\} \\
    & & & \sum_{i=1}^{N} E_{dur}(d_i; \theta^*) \cdot x_{ij} = E_{j}, & \forall j \in \{1, \dots, m\}  \nonumber  \\
    % & & & \sum_{i=1}^{N} L_{{dur}_i} \cdot x_{ij} = L_{j}, & \forall j \in \{1, \dots, N_{mb}^*\} \\
    & & & \sum_{i=1}^{N} L_{dur}(d_i; \theta^*) \cdot x_{ij} = L_{j}, & \forall j \in \{1, \dots, m\}  \nonumber \\
    & & & x_{ij} \in \{0, 1\}, & \forall i, j  \nonumber 
    % \end{aligned}
\end{align}
% \vspace{8pt}
}

In this formulation, the binary decision variable $x_{ij}$ is 1 if data item $d_i$ is assigned to microbatch $j$, and 0 otherwise. The first constraint ensures that each data item is assigned to exactly one microbatch. The subsequent two constraints define the total execution time for the modality encoder stage ($E_j$) and the LLM stage ($L_j$) for each microbatch $j$. These are calculated by summing the durations ($E_{dur}(d_i; \theta^*)$ and $L_{dur}(d_i; \theta^*)$) of the data items assigned to that microbatch. The objective function then seeks to minimize the maximum of these stage durations across all microbatches.

% \rs{the formulation doesn't seem to include communication overheads. if that is not required, then mention that point.}

% The core task of the Online Microbatch Scheduler is to solve a load-balancing problem for each incoming global batch. Given a global batch consisting of N data items and the number of microbatches m determined by the offline optimizer, the goal is to partition these N items into m distinct microbatches.

% First, for each data item i in the global batch, we use the performance models from the Profiling Engine to predict its execution duration on the vision encoder stage ($V\_dur_{i}$) and the LLM stage ($L\_dur_{i}$). The total execution time for a specific microbatch $j$ is then the sum of the durations of the data items assigned to it. Since the pipeline stages operate in parallel, the overall pipeline throughput is dictated by the maximum processing time encountered across any stage of any microbatch.

% Let $C_{max}$ represent this maximum processing time, which acts as the bottleneck for the entire pipeline execution for the current global batch. The objective is to find a partition of data into microbatches that minimizes this $C_{max}$. This can be formulated as a min-max optimization problem. We seek to find the partition that minimizes the maximum cumulative duration among all stages of all m microbatches. The problem is subject to the constraint that each data item must be assigned to exactly one microbatch.

\subsubsection{Scheduling Optimization}
\label{sec:sched_opt}
\begin{figure}[t]
    \centering
    \includegraphics[width=0.9\columnwidth,trim={0pt 0pt 0pt 0pt},clip]{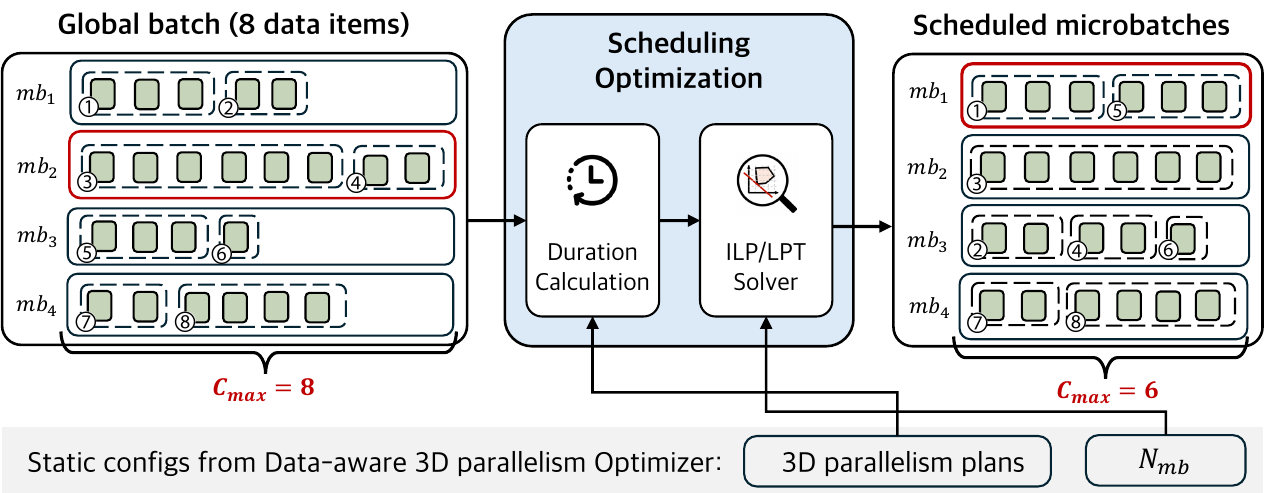}
    \caption{The scheduling optimization process dynamically partitions each global batch with $N$ data items into $N_{mb}$ microbatches, when $N=8$ and $ N_{mb}=4$.}
    % \revisionCommon{}{The scheduling optimization process dynamically partitions each global batch with $N$ data items into $N_{mb}$ microbatches, when $N=8$ and $ N_{mb}=4$.}}
    \label{fig:scheduler}
    % \vspace{-20pt}
\end{figure}

\revisionnoteCommon{R1.O3}\revisionCommon{}{The process for solving the load }\revisionnoteCommon{R2.O5}\revisionCommon{}{balancing problem is detailed in \mbox{\autoref{fig:scheduler}}. The scheduler operates }\revisionnoteCommon{R2.O6}\revisionCommon{}{asynchronously to eliminate scheduling overhead. While the model }\revisionnoteCommon{R2.O7}\revisionCommon{}{executes the computation for the current iteration, the scheduler processes the subsequent global batch in parallel on the CPU. The scheduler begins by receiving the static 3D parallelism plan and $N_{mb}$ from the Data-aware 3D Parallelism Optimizer. When a new global batch of $N$ data items arrives, the scheduler first calculates the execution duration for each item. This calculation is derived by dividing the computational load of each item by its expected throughput under the active 3D parallel plan.

These per item duration values serve as inputs to a hybrid solving mechanism. To ensure low runtime latency, we first employ an ILP solver with a strict time limit to search for a solution that minimizes the objective function $C_{max}$. If the solver fails to converge within the allocated time, the scheduler reverts to the Longest Processing Time (LPT) heuristic. This greedy approach sorts the data items in descending order of duration and iteratively assigns each item to the microbatch with the lowest current accumulated load. Upon determining a valid assignment through either method, the mechanism returns a set of index groups that defines the target partition of the global batch. Based on these indices, the scheduler partitions the data into scheduled microbatches.}
% The process for solving the load-balancing problem is detailed in \autoref{fig:scheduler}. The scheduler operates asynchronously to eliminate scheduling overhead. While the model executes the computation for the current iteration, the scheduler processes the subsequent global batch in parallel on the CPU. The scheduler begins by receiving the optimal 3D parallelism plan and the number of microbatches ($m$) from the Data-aware 3D Parallelism Optimizer. When a new global batch of $N$ data items arrives, the scheduler first predicts the execution duration for each item. This prediction is derived by dividing the calculated computational load of each item by its expected throughput under the active 3D parallel plan. These per-item duration tuples, representing the expected execution time on both the modality encoder and the LLM, are then passed to the ILP solver. The solver finds an assignment of data items to microbatches that minimizes the objective function $C_{max}$, as formulated in the previous section. Upon finding a solution, the solver returns a set of index groups that defines the optimal partition of the global batch. Based on these indices, the scheduler partitions the data into scheduled microbatches.

\mypar{Scheduling Optimization Time Complexity Analysis} 
\revisionnote{R2.O5}\revision{}{The complexity }\revisionnote{R2.O6}\revision{}{of the hybrid scheduling mechanism derives from its two }\revisionnote{R2.O7}\revision{}{components. The ILP formulation assigns each of the $GBS$ items to one of $m$ buckets, resulting in an exponential worst-case complexity of $O(m^{GBS})$. Despite this theoretical bound, the solver remains efficient by leveraging branch pruning to reduce the search space ~\cite{gurobi, ortools}, while the scheduler utilizes asynchronous execution to hide solving latency. The fallback LPT heuristic incurs polynomial complexity; by iteratively assigning items to the bucket with the lowest accumulated load, it achieves a complexity of $O(GBS \cdot \log m)$ ~\cite{LPT_Graham_1969}.}

\subsubsection{Adaptive Correction} 
\label{sec:adaptive_corr}
\revisionnote{R1.O2}\revision{}{Under stable operating conditions, stage throughput is largely predictable as a function of input shape using profiling-based interpolation. However, modern GPU execution stacks frequently invoke different specialized kernels or optimized execution strategies depending on the specific dimensions of the input tensor~\cite{FlashDecodingpp_MLSYS24}. This behavior leads to non-smooth and regime-dependent performance characteristics. Consequently, a small subset of rare input shapes may exhibit throughput behavior that consistently deviates from interpolation-based predictions.

The Adaptive Correction mechanism addresses these deviations by continuously tracking the execution metrics. The system quantifies the expected benefit $B$ for a given input shape as the difference between the actual and predicted throughput:}
\begin{equation}
    B = Th_{actual} - Th_{pred}
\end{equation}
\revisionnote{R1.O2}\revision{}{where $Th_{actual}$ and $Th_{pred}$ denote the measured runtime throughput and the interpolated prediction, respectively. This deviation translates to potential makespan degradation in the worst-case scenario. To mitigate this, the mechanism identifies such edge cases and provides feedback to the scheduling solver by updating a penalty function with the observed data. The scheduling solver utilizes this adjusted information when partitioning future global batches to ensure accurate load balancing.

To manage the overhead associated with continuous monitoring, we employ a cost-benefit analysis. The cost $C$ is quantified during the initial training phase by measuring the throughput differential when the tracking mechanism is first deactivated and subsequently activated. The system evaluates the average benefit $B$ over a sequence of $I$ iterations. If this average $B$ fails to exceed the recurring cost $C$, the tracking mechanism is deactivated, allowing the training process to proceed without this monitoring overhead to preserve overall efficiency.}
\section{Implementation}
\label{sec:impl}
\begin{figure}[t!]
    \centering
    \includegraphics[width=0.8\columnwidth,trim={0pt 0pt 0pt 0pt},clip]{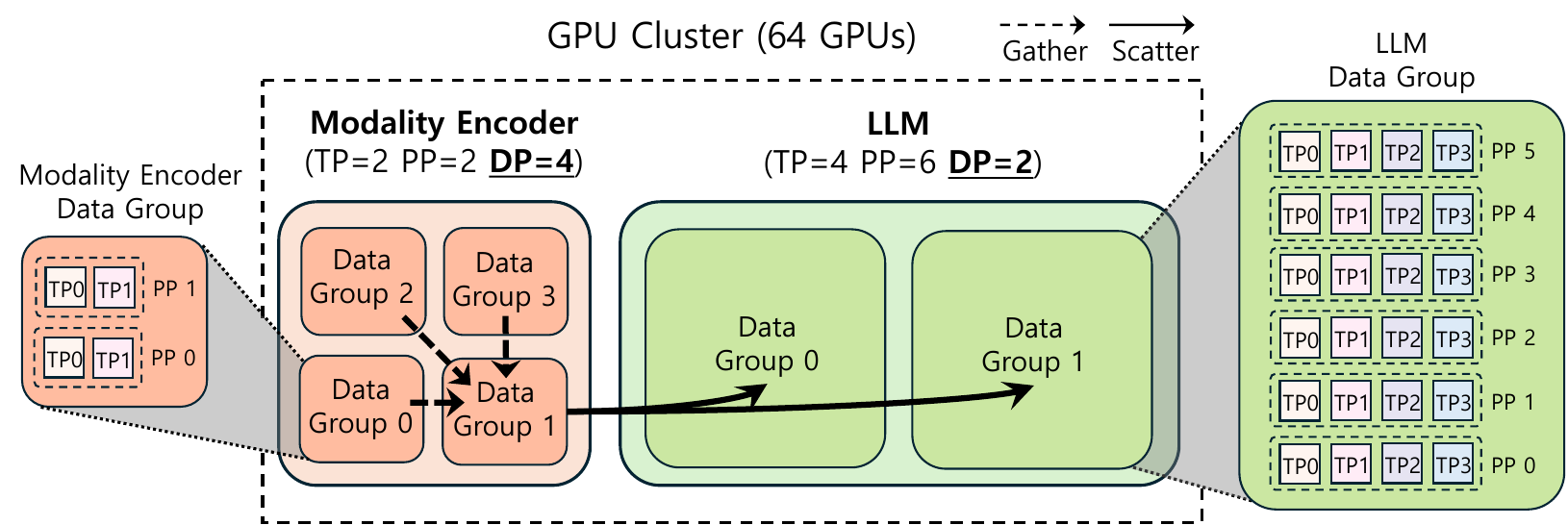}
    \caption{The Inter-model Communicator resolves data group mismatches in heterogeneous configurations. In the forward pass, activations from the modality encoder are gathered and scattered to the LLM data groups. This operation reverses during the backward pass to propagate gradients from the LLM back to the modality encoder.}
    \label{fig:inter_model_comm}
\end{figure}

The system is implemented on top of PyTorch~\cite{pytorch} as it offers the necessary flexibility to leverage native distributed communication primitives and parallelism functionalities. The core contribution lies in the development of a mechanism that enables independent 3D parallelism for distinct model components, a feature not supported by standard frameworks.

\revisionnote{R2.O9}\revision{}{\mbox{\sysname} supports heterogeneous 3D parallel configurations for the modality encoder and the LLM. This flexibility contrasts with conventional distributed training frameworks that enforce identical TP and DP degrees across the entire model, which necessitates static alignment of data groups and restricts module-specific optimization. To eliminate these constraints, the available GPUs are partitioned into distinct process groups for each module. While allowing distinct data groups expands the search space for efficient configurations, it introduces a communication mismatch when the data groups of the two modules differ in size.

% \autoref{fig:inter_model_comm} illustrates the inter-model communicator mechanism introduced to resolve this communication mismatch. Within the data group of the modality encoder, a single predetermined rank, representing a specific GPU process, is designated to assume the role of the communicator. During the forward pass, this rank is responsible for gathering activations from all ranks within its local data group and subsequently scattering these activations to the corresponding ranks in the data group of the LLM. Conversely, during the backward pass, the communicator gathers gradients from the LLM group and scatters them back to the modality encoder group.
As illustrated in \mbox{\autoref{fig:inter_model_comm}}, we introduce an inter-model communicator mechanism to resolve this disparity. Conventional frameworks cannot facilitate communication in the scenario shown in the figure, where the modality encoder is configured with $DP=4$ and the LLM with $DP=2$, due to the group size mismatch. \mbox{\sysname} overcomes this by designating a single predetermined rank within the modality encoder's data group as the communicator. Specifically, the communicator gathers output activations from the four data groups of the modality encoder and subsequently scatters them to the two corresponding data groups of the LLM during the forward pass. Conversely, during the backward pass, the mechanism operates in reverse, gathering gradients from the LLM groups and scattering them back to the modality encoder groups to ensure consistency.}

\revisionnote{R1.M1}\revision{}{Our decoupled design generalizes to frameworks supporting native 3D parallelism and flexible process group management, such as PyTorch and Megatron-LM. However, \mbox{\sysname} is currently incompatible with DeepSpeed, as it does not natively support 3D parallelism and lacks the flexible pipeline parallelism primitives required for independent module execution.}
\section{Evaluation}
\label{sec:evalutaion}

\begin{table}[t!]
\setlength{\textfloatsep}{10pt plus 2pt minus 1pt}
\centering
\caption{Composition of the mixed dataset.}
\small
\label{tab:dataset_composition}
\begin{tabular}{ccc} % 순서 변경: l (Dataset Name), r (Data Type), l (\# of Samples)
\toprule % \hline 대신 \toprule 사용 (booktabs)
\textbf{Dataset Name}   & \textbf{Data Type}       & \textbf{\# of Samples} \\
\midrule % \hline 대신 \midrule 사용
LLaVA-Wild  \cite{llava}           & \multirow{3}{*}{Single Image} & 28k                    \\
AI2D  \cite{AI2D}                  &                               & 18k                    \\
Infographic VQA  \cite{Infographicvqa}       &                               & 19k                    \\
\midrule % \hline 대신 \midrule 사용
M4-Instruct \cite{m4-instruct}            & Multiple Images               & 60k                    \\
\midrule % \hline 대신 \midrule 사용
LLaVA-Video \cite{llava-video}            & Video                         & 60k                    \\
\bottomrule % \hline 대신 \bottomrule 사용
\end{tabular}
\vspace{-5pt}
\end{table}

\begin{table}[t!]
\centering
\small
\caption{Model configurations of the evaluated MLLMs.}
\label{tab:mllm_configs}
\begin{tabular}{ccc}
\toprule % \hline 대신 \toprule 사용 (booktabs)
\textbf{\makecell{MLLM \\ Architecture}} & \textbf{Modality Encoder} & \textbf{LLM Backbone} \\
\midrule % \hline 대신 \midrule 사용
\multirow{2}{*}{LLaVA-OV \cite{llavaov}} & \multirow{2}{*}{SigLIP \cite{siglib}} & Qwen-2.5 \cite{qwen} (7B, 32B, 72B) \\
                          &                         & Llama-3 \cite{llama3} (8B, 70B)       \\
\midrule % \hline 대신 \midrule 사용
InternVL-2.5 \cite{internvl}              & InternViT \cite{internvl}              & Qwen-2.5 \cite{qwen}  (72B)        \\
\bottomrule % \hline 대신 \bottomrule 사용
\end{tabular}
\vspace{-10pt}
\end{table}

In this section, we present a comprehensive evaluation of \sysname to demonstrate its effectiveness in optimizing MLLM training. We begin by detailing our experimental setup, outlining the hardware testbed, the baseline systems used for comparison, and the models and datasets selected for our experiments (\autoref{sec:exp_setup}). The evaluation then proceeds with a series of macro-experiments that measure the end-to-end performance of \sysname against these baselines across a variety of MLLM configurations (\autoref{sec:mac}). To provide a deeper understanding of the underlying sources of the observed performance gains, we subsequently conduct a series of micro-experiments (\autoref{sec:micro}).
% evaluation summary end to end로 얼마 성능 보였다.
% micro evaluation 에서는 scale 에서는 얼마
% bubble 이랑 stage wise throughput 관련된 내용
% mypar 부분은 지워도 될듯 
% 1 to 32 nodes
Our key findings are summarized as follows:

\begin{itemize} [left=0pt]
\item \sysname achieves up to 3.6{\small$\times$} higher training throughput compared to PyTorch and Megatron-LM across various MLLM architectures and model scales, demonstrating its effectiveness in optimizing distributed training efficiency.

\item The performance gap between \sysname and baseline systems widens as the number of nodes increases, indicating that \sysname scales more effectively in larger cluster environments.
% \item As the system scales from 8 to 64 GPUs, \sysname maintains near-linear throughput growth and continues to widen its performance gap over the baselines, confirming that its data-aware 3D parallelism optimization remains effective at larger cluster sizes.

\item Fine-grained micro-level analysis shows that \sysname reduces pipeline idle time by up to 84\% while simultaneously increasing stage-wise throughput and maintaining balanced performance across all pipeline stages, leading to significantly improved GPU utilization. This demonstrates that our approach effectively resolves the dual challenges of load imbalance and performance variability outlined in \autoref{sec:motivation}.
\end{itemize}

\vspace{-10pt}

% \begin{figure*}[th!]
%   \centering
%   \begin{subfigure}{0.99\textwidth}
%     \centering
%     \includegraphics[width=\linewidth,trim={10pt 10pt 10pt 9pt},clip]{figures/end_to_end.pdf}
%     \subcaption{End-to-end training throughput.}
%     \label{fig:train_thr}
%   \end{subfigure}
%   \begin{subfigure}{0.99\textwidth}
%     \centering
%     \includegraphics[width=\linewidth,trim={10pt 0pt 10pt 8pt},clip]{figures/training_time.pdf}
%     \subcaption{End-to-end training time.}
%     % \vspace{7mm}
%     \label{fig:train_time}
%   \end{subfigure}
%   \hfill
%   % \vspace{-5pt}
%   \caption{End-to-end training performance of \sysname over baseline systems.}
%   \label{fig:end_to_end}
%   % \vspace{-5pt}
% \end{figure*}

\subsection{Experimental Setup}
\label{sec:exp_setup}

% \begin{table}[t!]
% \centering
% \small
% \caption{Model configurations of the evaluated MLLMs.}
% \label{tab:mllm_configs}
% \begin{tabular}{ccc}
% \toprule % \hline 대신 \toprule 사용 (booktabs)
% \textbf{\makecell{MLLM \\ Architecture}} & \textbf{Modality Encoder} & \textbf{LLM Backbone} \\
% \midrule % \hline 대신 \midrule 사용
% \multirow{2}{*}{LLaVA-OV \cite{llavaov}} & \multirow{2}{*}{SigLIP \cite{siglib}} & Qwen-2.5 \cite{qwen} (7B, 32B, 72B) \\
%                           &                         & Llama-3 \cite{llama3} (8B, 70B)       \\
% \midrule % \hline 대신 \midrule 사용
% InternVL-2.5 \cite{internvl}              & InternViT \cite{internvl}              & Qwen-2.5 \cite{qwen}  (72B)        \\
% \bottomrule % \hline 대신 \bottomrule 사용
% \end{tabular}
% \end{table}

% \begin{figure*}[th!]
%   \centering
%   \begin{subfigure}{0.99\textwidth}
%     \centering
%     \includegraphics[width=\linewidth,trim={10pt 10pt 10pt 9pt},clip]{figures/end_to_end.pdf}
%     \subcaption{End-to-end training throughput.}
%     \label{fig:train_thr}
%   \end{subfigure}
%   \begin{subfigure}{0.99\textwidth}
%     \centering
%     \includegraphics[width=\linewidth,trim={10pt 0pt 10pt 8pt},clip]{figures/training_time.pdf}
%     \subcaption{End-to-end training time.}
%     % \vspace{7mm}
%     \label{fig:train_time}
%   \end{subfigure}
%   \hfill
%   % \vspace{-5pt}
%   \caption{End-to-end training performance of \sysname over baseline systems.}
%   \label{fig:end_to_end}
%   % \vspace{-5pt}
% \end{figure*}

The experiments evaluate the end-to-end training efficiency of \sysname against state-of-the-art baselines. To ensure a comprehensive assessment, the evaluation employs representative MLLM architectures and diverse datasets that mirror the workload heterogeneity found in real-world scenarios.
% In this section, we present the end-to-end performance of \sysname, comparing its throughput and training time against baseline systems. These experiments are designed to demonstrate the overall efficiency gains achieved by our integrated optimization approach across representative MLLM architectures and scales. Furthermore, we analyze the relationship between the computational load distribution within the MLLM and the resulting system performance, providing insights into the effectiveness of our data-driven methodology under varying model characteristics.

\mypar{Baselines} We compare our system against two strong baselines: Megatron-LM \cite{megatron-lm} and a custom PyTorch-based implementation. Megatron-LM is widely regarded as a state-of-the-art framework for large-scale model training, providing a highly optimized implementation of 3D parallelism. As our system is built upon PyTorch \cite{pytorch}, we also developed a baseline that leverages PyTorch's native distributed functionalities. For both baselines, the 3D parallelism configurations were manually tuned following conventional best practices to achieve their best possible performance \cite{megatron-lm}.
% \rs{Comparison with other works, e.g., Optimus (https://www.usenix.org/system/files/atc25-feng.pdf); DISTMM (https://www.usenix.org/system/files/nsdi24-huang.pdf)?}

\mypar{Hardware} Our experiments were conducted on a cluster of up to 8 nodes, each equipped with an NVIDIA HGX A100 8-GPU system. The GPUs within each node are interconnected via high-bandwidth NVLink, and the nodes are connected with an 800 Gbps InfiniBand network.

\mypar{Datasets} To create a realistic MLLM training scenario characterized by workload variability, we constructed a composite dataset from several public sources. This dataset mirrors the heterogeneity commonly found in MLLM pre-training and instruction tuning by including a mix of data types: single images, multiple images, and videos. The specific composition of the dataset is detailed in \autoref{tab:dataset_composition}.

\mypar{Models} We selected two state-of-the-art, publicly available MLLMs for our evaluation: LLaVA-OneVision (LLaVA-OV) \cite{llavaov} and InternVL-2.5 \cite{internvl}. To demonstrate the general applicability and robustness of our system across various architectures and scales, we configured these models with different encoders and LLM backbones. Specifically, our experiments utilized powerful modality encoders including SigLIP \cite{siglib} and InternViT \cite{internvl}. These were paired with a range of LLM backbones. The specific model configurations used in our experiments are detailed in \autoref{tab:mllm_configs}.

\subsection{Macro Experiments}
\label{sec:mac}
This section presents the end-to-end performance of \sysname compared against baseline systems. The experiments are designed to demonstrate the overall training efficiency gains achieved by the integrated optimization approach across a range of representative MLLM architectures and scales.
% In this section, we present the end-to-end performance of \sysname and compare it against baseline systems. These experiments are designed to demonstrate the overall throughput improvements achieved by our integrated optimization approach across a range of representative MLLM architectures and scales.

\subsubsection{End-to-end Performance}

\begin{figure*}[t!]
  \centering
  \begin{subfigure}{0.99\textwidth}
    \centering
    \includegraphics[width=\linewidth,trim={10pt 10pt 10pt 9pt},clip]{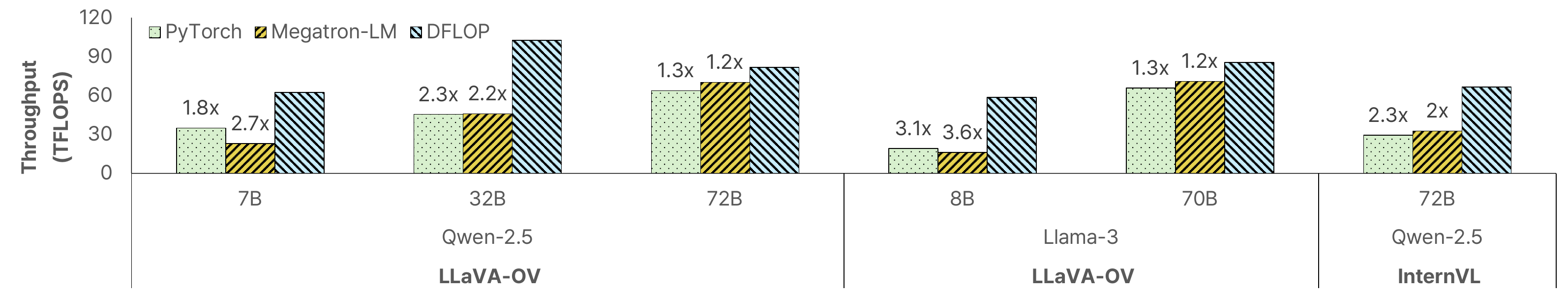}
    \subcaption{End-to-end training throughput.}
    \label{fig:train_thr}
  \end{subfigure}
  \begin{subfigure}{0.99\textwidth}
    \centering
    \includegraphics[width=\linewidth,trim={10pt 0pt 10pt 8pt},clip]{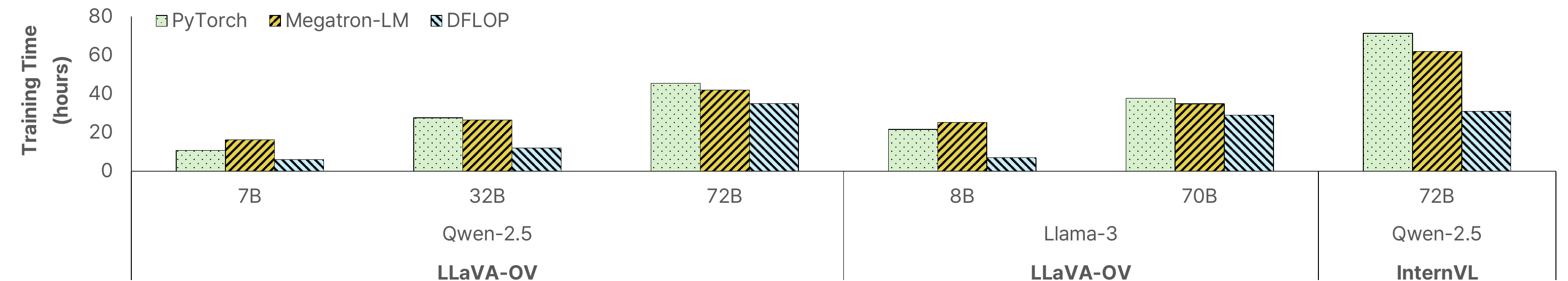}
    \subcaption{End-to-end training time.}
    % \vspace{7mm}
    \label{fig:train_time}
  \end{subfigure}
  \hfill
  % \vspace{-5pt}
  \caption{End-to-end training performance of \sysname over baseline systems.}
  \label{fig:end_to_end}
  % \vspace{-5pt}
\end{figure*}

In this section, we present the end-to-end performance of \sysname, comparing its training throughput and total training time against the baseline systems. \sysname consistently demonstrates substantial performance improvements. As shown in \autoref{fig:train_thr}, GPU throughput gains range from 1.2{\small$\times$} to 3.6{\small$\times$}, and \autoref{fig:train_time} illustrates that this translates to a reduction in total training time by 5 to 40 hours across various MLLM configurations.

The throughput results highlight the robustness of our approach. For the LLaVA-OV architecture, \sysname maintains its high performance advantage over the baselines even when the internal LLM backbone is varied in both architecture and scale. Furthermore, the significant gains observed with the InternVL architecture confirm that our system's effectiveness is not limited to a single MLLM design but generalizes across different architectural paradigms. In terms of practical impact, we applied the official training recipes for LLaVA-OV and InternVL to measure the total training time. The results confirm that the observed throughput gains translate directly into significant real-world time savings.

% This performance advantage stems from the fundamental difference in optimization methodology. The baseline systems are data-agnostic and apply a single, monolithic 3D parallelism strategy to the entire MLLM, failing to account for the heterogeneous characteristics of the internal modules or the data workload. In contrast, \sysname orchestrates three distinct components. The Profiling Engine first gathers detailed performance models and dataset statistics. Based on this information, the Data-aware 3D Parallelism Optimizer selects an static configuration that minimizes the expected makespan for the specific workload. Subsequently, at runtime, the Online Microbatch Scheduler dynamically minimizes pipeline bubbles for each iteration.
This performance advantage stems from the fundamental difference in optimization methodology. Unlike baseline systems that enforce a uniform, data-agnostic strategy, \sysname adopts a holistic approach that aligns system resources with the specific computational demands of each module. By leveraging comprehensive profiling, the system derives heterogeneous parallel configurations that minimize the expected makespan based on dataset statistics. This static optimization is bridged with runtime execution through dynamic scheduling, which actively mitigates pipeline bubbles caused by input variability.

\begin{figure}[t!]
    \centering
    \includegraphics[width=0.8\linewidth,trim={10pt 7pt 10pt 5pt},clip]{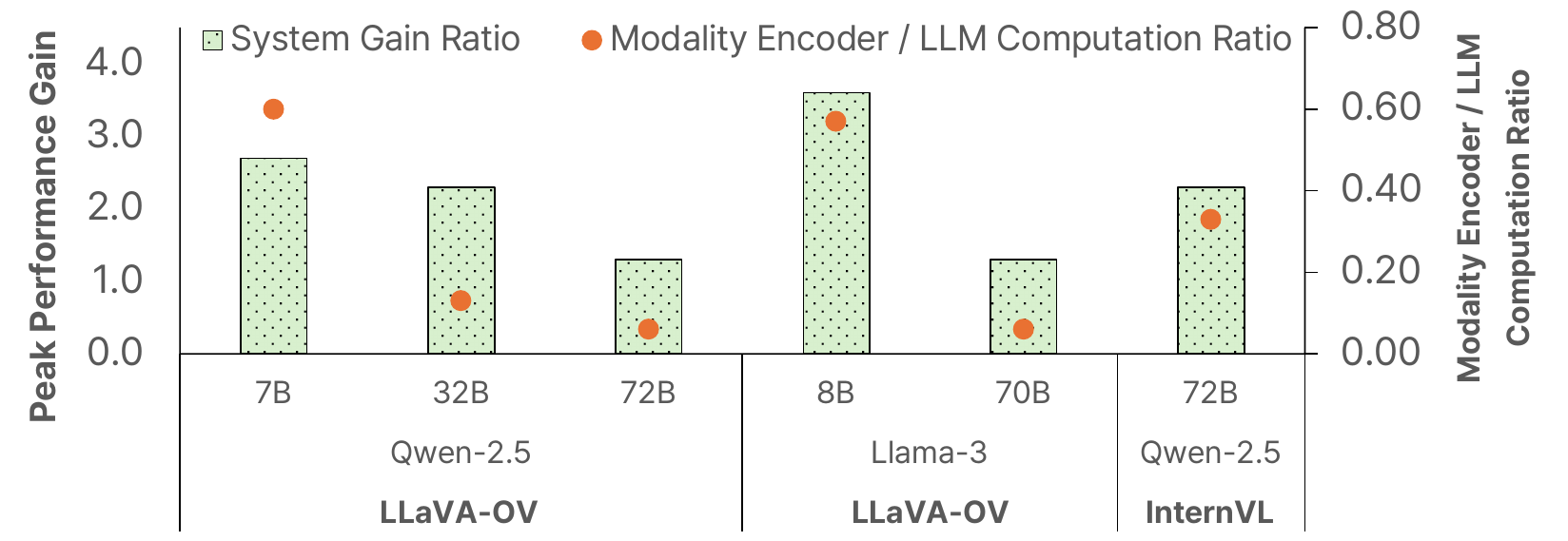}
    %\vspace{-4mm}
    \caption{Correlation between the computational load ratio (modality encoder FLOP / LLM FLOP) and the maximum performance gain achieved by \sysname when compared to baseline systems, shown for various MLLM configurations.}
    \label{fig:compute_ratio}
    \vspace{-10pt}
\end{figure}

\subsubsection{Impact of Computational Asymmetry}
\label{sec:impact_asym}
This section evaluates the impact of the computational ratio between the modality encoder and the LLM on the performance gain of \sysname. \autoref{fig:compute_ratio} demonstrates that the performance advantage of \sysname amplifies as the computational loads between the two modules become more balanced. This trend stems from the structural limitations of baseline systems, which enforce homogeneous parallel configurations across the entire model. As the computational distribution becomes symmetric, this monolithic constraint precludes independent optimization of heterogeneous modules, leading to significant resource underutilization. In contrast, \sysname maximizes end-to-end throughput by decoupling parallel strategies for each module and dynamically mitigating workload skew, thereby widening the performance gap against baselines in balanced workloads.

\revisionnote{R3.O1}\revision{}{This capability is particularly significant given the current trajectory of MLLM development toward balanced cross-modal computation. This shift is driven by the emergence of larger modality encoders~\cite{qwenvl, internvl} and more efficient LLM backbones~\cite{deekseekvl, llama4}, alongside strategies reducing the number of tokens processed by the LLM~\cite{qwenvl, internvl}. Consequently, the utility of \mbox{\sysname} is expected to grow as future MLLM architectures converge toward equiproportional computation.}

\subsection{Micro Experiments}
\label{sec:micro}
To provide a deeper analysis of the end-to-end performance gains and their underlying sources, we conduct a series of targeted experiments, each designed to examine a specific aspect of \sysname's design and behavior. 
\subsubsection{Generalization Across Modalities}
\label{sec:more_modal}
\begin{figure}[t!]
    \centering
    \includegraphics[width=0.8\columnwidth,trim={15pt 12pt 5pt 10pt},clip]{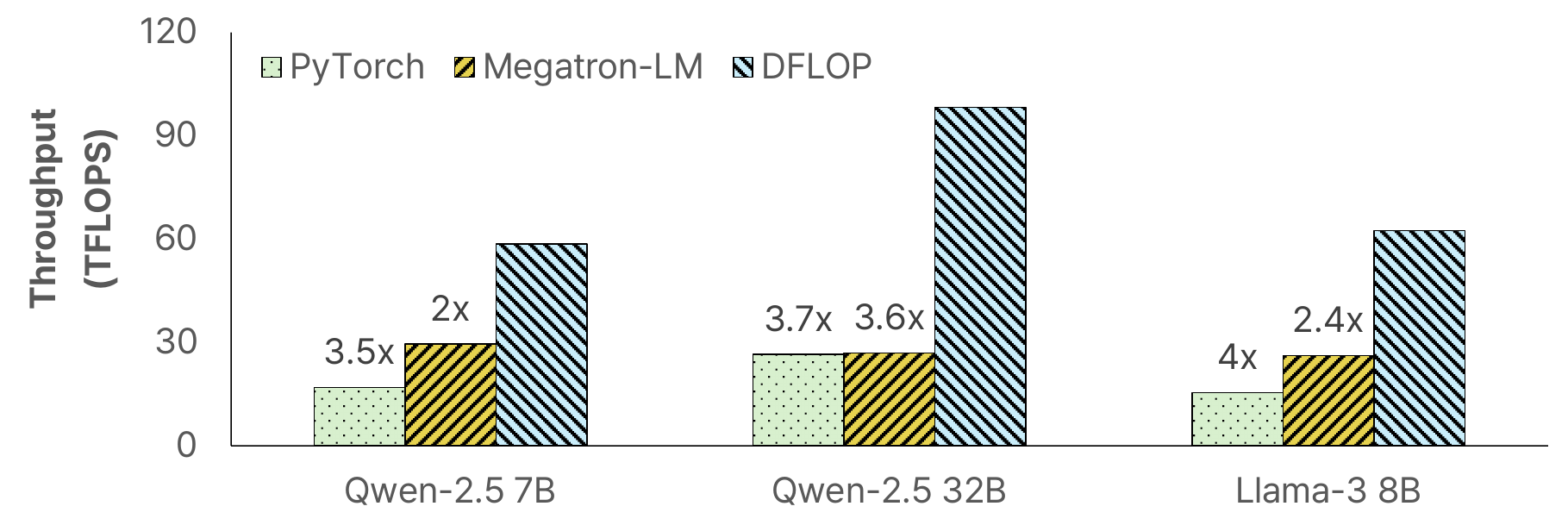}
    \caption{Performance gain demonstrating the cross-modal generalization of \sysname, measured on a 4-node cluster with 8×A100 GPUs per node. Using the Qwen2-Audio architecture as a representative case.}% 
    \label{fig:audio_modality}
    % \vspace{-10pt}
\end{figure}

\begin{figure}[t!]
    \centering
    \includegraphics[width=0.8\columnwidth,trim={15pt 10pt 5pt 6pt},clip]{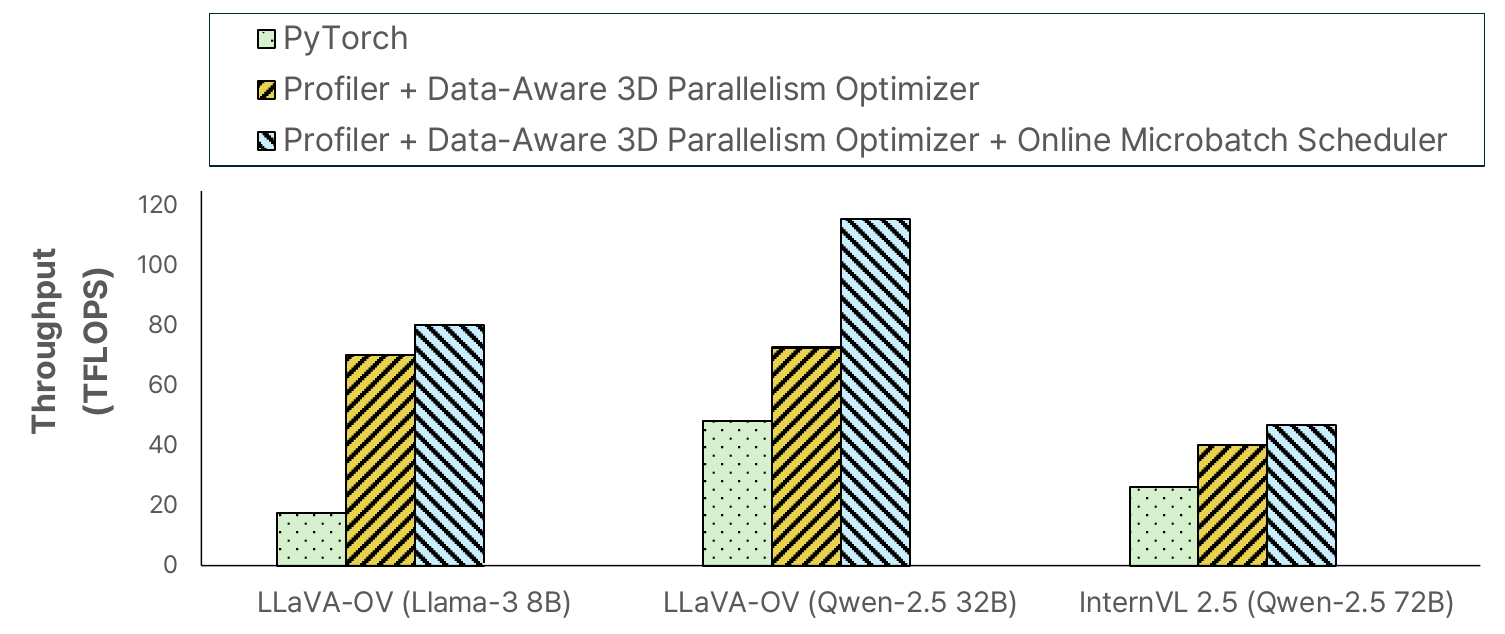}
    \caption{Performance gain on the mixed dataset using a 4-node cluster with 8×A100 GPUs per node, obtained by incrementally adding DFLOP components to the PyTorch baseline across various MLLM configurations.}% \kp{legend is too small}}
    \label{fig:ablation_study}
    % \vspace{-10pt}
\end{figure}

To evaluate the extensibility of \mbox{\sysname} beyond vision-language models, \revisionnote{R1.O1}\revision{}{we conduct experiments on the Qwen2-Audio~\cite{qwen_audio} architecture, a representative audio-language MLLM. As illustrated in \mbox{\autoref{fig:audio_modality}}, \mbox{\sysname} achieves significant throughput gains ranging from 2{\small$\times$} to 4{\small$\times$} for the audio modality as well. These substantial improvements are attributed to the balanced computational intensity between the audio modality encoder and the LLM. Specifically, while the audio modality encoder involves high computational demands, it utilizes an average pooling layer at the end of its pipeline to reduce the number of tokens fed into the LLM. This architectural feature decreases the LLM workload, resulting in a comparable distribution of compute requirements between the encoder and the LLM.}

\begin{figure*}[t!]
  \centering
  \begin{subfigure}{0.5\textwidth}
    \centering
    \includegraphics[width=\linewidth,trim={10pt 9pt 10pt 6pt},clip]{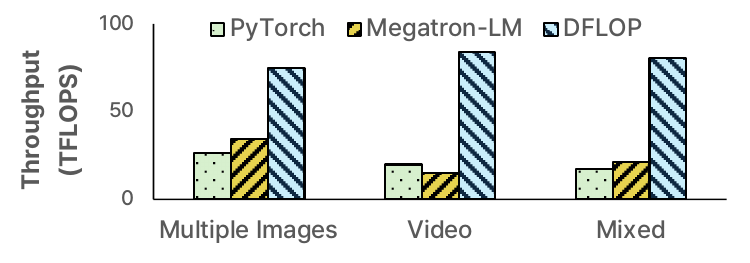}
    \subcaption{Performance on diverse datasets.}
    \vspace{10pt}
    \label{fig:diverse_dataset}
  \end{subfigure}
  \begin{subfigure}{0.99\textwidth}
    \centering
    \includegraphics[width=\linewidth,trim={10pt 10pt 10pt 7pt},clip]{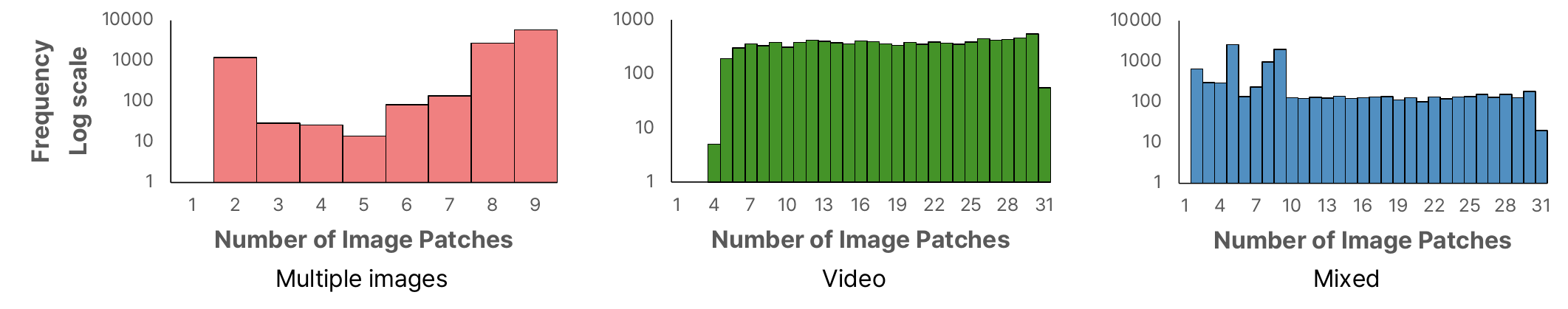}
    \subcaption{Input frequency distribution by the number of image patches.} %\kp{need to specify what x axis means.}}
    % \vspace{7mm}
    \label{fig:data_dist}
  \end{subfigure}
  \hfill
  \caption{System performance of LLaVA-OV (Llama-3 8B) on a 4-node cluster with 8xA100 GPUs per node and input characteristics across diverse datasets. The mixed dataset integrates single-image, multiple-image, and video inputs, revealing the degree of data heterogeneity in each dataset.}
  \label{fig:patch-triple}
  \vspace{-5pt}
\end{figure*}

\subsubsection{Ablation Study}
\label{sec:ablation_study}
% \begin{table}[t!]
% \centering
% \caption{Ablation study of throughput (samples/sec) for different optimization methods.}
% \label{tab:ablation_study}
% \scalebox{0.7}{
%     \begin{tabular}{llcc}
%     \toprule
%     \textbf{Models} & \textbf{Method} & \textbf{Throughput} \\
%     \midrule
%     \multirow{3}{*}{LLaVA-OV (Llama-3 8B)} 
%      & PyTorch Baseline & 38.84 \\
%      & + Data-Aware 3D Parallelism Optimization & 62.22 (1.6x)\\
%      & + Online Microbatch Scheduling & 85.57 (2.2x) \\
%     \midrule
%     \multirow{3}{*}{InternVL (Qwen-2.5 72B)} 
%      & PyTorch Baseline & 38.84 \\
%      & + Data-Aware 3D Parallelism Optimization & 62.22 (1.6x) \\
%      & + Online Microbatch Scheduling & 85.57 (2.2x) \\
%     \bottomrule
%     \end{tabular}
%     }
% \end{table}
The individual contributions of the Data-aware 3D Parallelism Optimizer and the Online Microbatch Scheduler are evaluated across diverse MLLM configurations, as shown in \autoref{fig:ablation_study}.

The primary source of performance gain shifts depending on the model configuration. For LLaVA-OV (Llama-3 8B), the Data-aware 3D Parallelism Optimizer delivers the majority of the improvement, whereas for LLaVA-OV (Qwen-2.5 32B), the Online Microbatch Scheduler becomes the dominant contributor. InternVL 2.5 (Qwen-2.5 72B) occupies a middle ground, with both components contributing comparably to the total gain. This variation correlates with the degree of computational asymmetry. In configurations with low computational asymmetry, such as LLaVA-OV (Llama-3 8B), the pipeline is highly sensitive to the static partitioning strategy, making the optimizer critical. Conversely, in highly asymmetric cases like LLaVA-OV (Qwen-2.5 32B), the dominant module dictates the pipeline bottleneck, diminishing the impact of static partitioning inefficiencies. Consequently, the optimization burden shifts to the scheduler to mitigate runtime bubbles caused by input variability.

\subsubsection{Performance on Diverse Datasets}
\label{sec:data_workload}
% \begin{figure}[t]
%     \centering
%     \captionsetup{labelfont={color=blue},textfont={color=blue}}
%     \begin{minipage}[t]{0.32\textwidth}
%     \includegraphics[width=\linewidth]{figures/Number_of_image_patches.png-1.pdf}
%     \caption{Mutiple images}
%     \label{fig:yfcc}
%     \end{minipage}\hfill%
%     \begin{minipage}[t]{0.32\textwidth}
%     \includegraphics[width=\linewidth]{figures/Number_of_image_patches.png-2.pdf}
%     \caption{Video}
%     \label{fig:multi-1-table}
%     \end{minipage}\hfill%
%     \begin{minipage}[t]{0.32\textwidth}
%     \includegraphics[width=\linewidth]{figures/Number_of_image_patches.png-3.pdf}
%     \caption{all}
%     \label{fig:multi-2-tables}
%     \end{minipage}
% \end{figure}

This experiment evaluates the robustness of \sysname across three distinct workload scenarios: multiple-image, video, and mixed datasets. As illustrated in \autoref{fig:diverse_dataset}, the throughput of PyTorch and Megatron-LM degrades as workload heterogeneity increases. While the limitations of these frameworks are less pronounced on the multiple-image dataset, their throughput declines on the heterogeneous video and mixed workloads. In contrast, \sysname maintains consistent high performance regardless of the dataset composition.

This performance differential is directly correlated with the input shape distributions presented in \autoref{fig:data_dist}. The multiple-image workload exhibits a narrow distribution concentrated within a limited range, resembling a static workload. In such homogeneous scenarios, the structural rigidity of static partitioning results in limited performance degradation compared to dynamic cases. However, video and mixed datasets introduce broad, uniform distributions with high per-iteration variance. These dynamic conditions expose the inflexibility of data-agnostic frameworks, which lack the mechanisms to adapt to fluctuating computational demands, leading to resource underutilization. Conversely, \sysname mitigates these variations through data-aware optimization and online scheduling, demonstrating its inherent resilience to the workload heterogeneity characteristic of modern MLLM training.

\subsubsection{GPU Cluster Scalability}
\label{sec:gpu_scale}

\begin{figure}[t!]
    \centering
    \includegraphics[width=0.75\columnwidth,trim={10pt 18pt 10pt 7pt},clip]{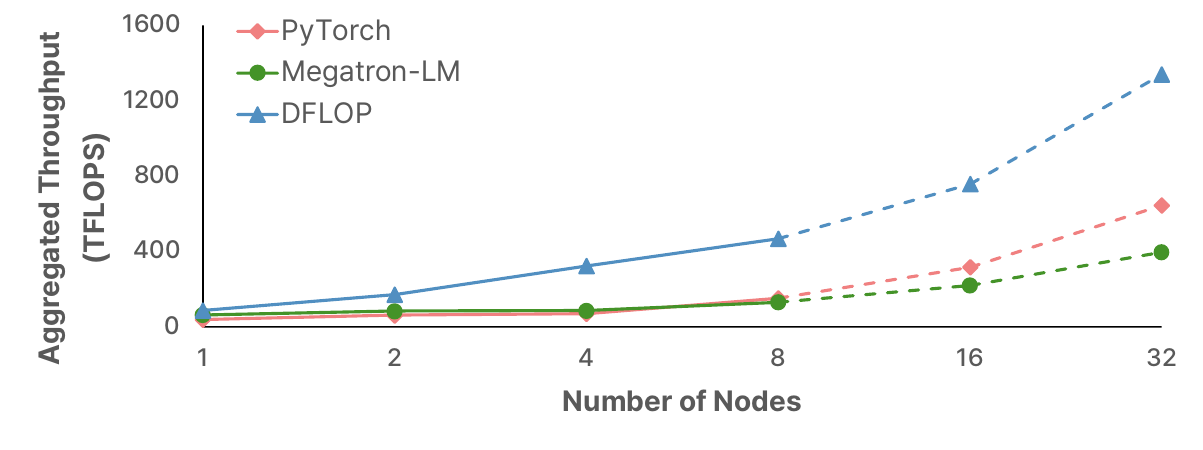}
    % \vspace{-12pt}
    \caption{Total cluster throughput variation of the LLaVA-OV (Llama-3 8B) on the mixed dataset, with increasing GPU node scale and 8×A100 GPUs per node. Measured results (1–8 GPU nodes) are shown with solid lines, and projected performance (16–32 GPU nodes) with dashed lines.}
    % \kp{DDMLLM to DFLOP}
    %\kp{Typo in Y axis}
    \label{fig:scalability}
    % \vspace{-15pt}
\end{figure}

This experiment evaluates the scalability of our system and projects its performance as the number of GPU nodes increases. The results, depicted in \autoref{fig:scalability}, illustrate the aggregate GPU throughput across different system scales. For this test, we used the LLaVA-OV model configured with a Llama-3 8B LLM. We measured the actual performance on clusters ranging from 1 to 8 nodes and then used these measurements to project the expected throughput for 16 and 32-node configurations.

The results clearly indicate that the performance gap between \sysname and the baseline systems not only persists but also widens as the cluster size grows. This superior scalability can be attributed to two primary factors, corresponding to the core components of our system. First, as more GPUs become available, the search space for the Data-aware 3D Parallelism Optimizer expands exponentially. This allows the optimizer to discover more fine-grained and effective parallelism strategies that are inaccessible at smaller scales, leading to greater performance improvements. Second, scaling up the number of GPUs often involves increasing the data parallelism degree, which in turn increases the number of data-parallel groups that must synchronize. In the baseline systems, the lack of a data-aware scheduling mechanism leads to significant overhead from straggler effects during these synchronization steps. In contrast, the Online Microbatch Scheduler in \sysname effectively mitigates this issue by balancing the workload across all data-parallel ranks, preventing stragglers and minimizing idle time. This effect becomes more pronounced at larger scales, further amplifying the performance advantage of our system.

\subsubsection{Analysis of Pipeline Bubble}
\label{sec:bubble}
% \begin{figure}[t]
%     \centering
%     \includegraphics[width=1\columnwidth,trim={10pt 6pt 10pt 12pt}, clip]{figures/pipeline_bubble.pdf}
%     \caption{GPU idle time due to pipeline bubbles on the mixed dataset for LLaVA-OV (Llama-3 8B) on a 4-node cluster with 8×A100 GPUs per node. The Ideal case represents the theoretical idle time from pipeline bubbles inherent to the 1F1B schedule for the given 3D parallel configuration, while the Real case represents the empirically measured idle time.}  
%     \label{fig:pipeline_bubble}
%     % \vspace{-15pt}
% \end{figure}

\begin{figure}[t]
    \centering
    \includegraphics[width=0.7\columnwidth,trim={3pt 3pt 10pt 8pt}, clip]{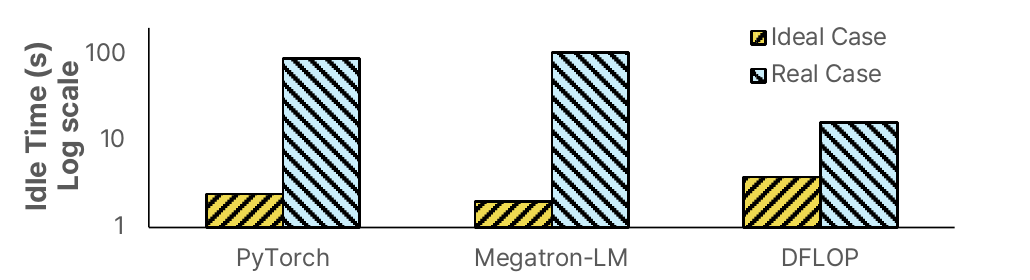}
    \caption{GPU idle time due to pipeline bubbles on the mixed dataset for LLaVA-OV (Llama-3 8B) on a 4-node cluster with 8×A100 GPUs per node. The Ideal case represents the theoretical idle time from pipeline bubbles inherent to the 1F1B schedule for the given 3D parallel configuration, while the Real case represents the empirically measured idle time.}  
    \label{fig:pipeline_bubble}
    % \vspace{-15pt}
\end{figure}

% \begin{figure}[t]
%     \centering
%     \includegraphics[width=1\columnwidth,trim={13pt 3pt 10pt 8pt}, clip]{figures/pipeline_bubble_8.pdf}
%     \caption{GPU idle time due to pipeline bubbles on the mixed dataset for LLaVA-OV (Llama-3 8B) on a 4-node cluster with 8×A100 GPUs per node. The Ideal case represents the theoretical idle time from pipeline bubbles inherent to the 1F1B schedule for the given 3D parallel configuration, while the Real case represents the empirically measured idle time.}  
%     \label{fig:pipeline_bubble}
%     % \vspace{-15pt}
% \end{figure}

As illustrated in \autoref{fig:pipeline_bubble}, \sysname reduces the empirically measured idle time by 82\% and 84\% compared to PyTorch and Megatron-LM, respectively. 
This result is particularly insightful because \sysname's theoretical minimum idle time, derived from the $(p-1)/m$ formula \cite{megatron-lm}, is slightly higher than that of the baselines. This apparent contradiction highlights a fundamental limitation in relying on idealized metrics for optimizing dynamic, real-world systems.

The theoretical minimum idle time is predicated on a perfectly homogeneous workload, a condition fundamentally violated in MLLM training, as illustrated in \autoref{fig:unbalance_pipe}. The baseline systems, implicitly optimized for this flawed ideal, are brittle to workload variations. This structural mismatch causes their empirically measured idle time to deviate substantially from their theoretical minimum.

In contrast, \sysname is explicitly designed for such heterogeneity. Its optimizer deliberately selects a smaller number of microbatches, a choice that is suboptimal from a purely theoretical standpoint but is strategically advantageous in a dynamic environment. This approach prioritizes real-world throughput and provides the Online Microbatch Scheduler with larger, more effective batches to balance. Consequently, \sysname's measured performance remains in close alignment with its theoretical potential. This demonstrates that the higher efficiency of \sysname is achieved not by optimizing for an unrealistic theoretical formula, but by robustly managing the real-world dynamics of the workload.

\subsubsection{Stage-wise Throughput}
\label{sec:stage_thr}
\begin{figure}[t]
    \centering
    \includegraphics[width=0.7\columnwidth,trim={10pt 10pt 15pt 10pt},clip]{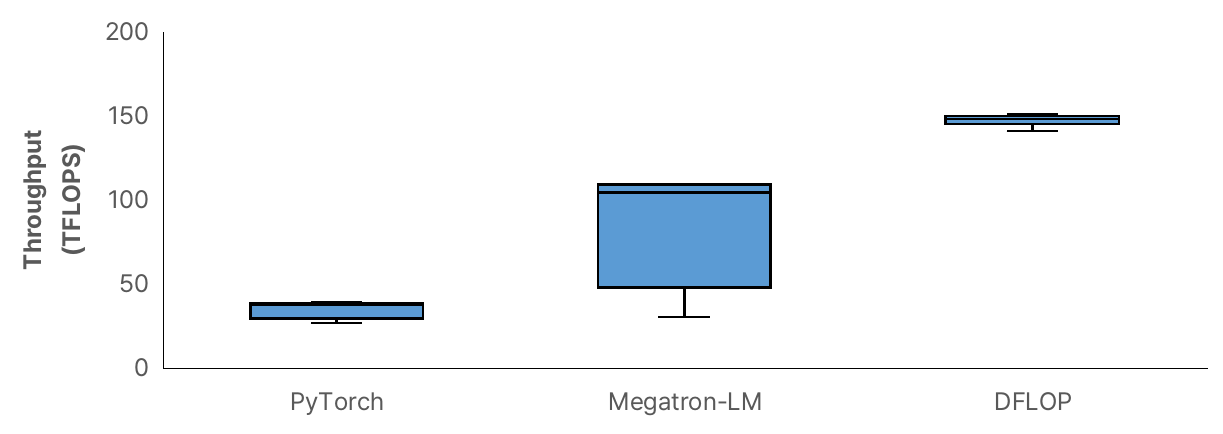}
    %\vspace{-4mm}
    \caption{Boxplots illustrating the distribution of throughput across pipeline stages on the mixed dataset for each evaluated system using LLaVA-OV (Llama-3 8B) on a 4-node cluster with 8×A100 GPUs per node.}
    % Stage-wise throughput for PyTorch, Megatron-LM, and DFLOP
    % Boxplots indicate 25th, median, and 75th percentile; cwhiskers indicate min and max values.
    \label{fig:stage_thr}
    % \vspace{-6pt}
\end{figure}

\begin{figure}[t]
    \centering
    \includegraphics[width=0.7\columnwidth,trim={15pt 12pt 15pt 5pt}, clip]{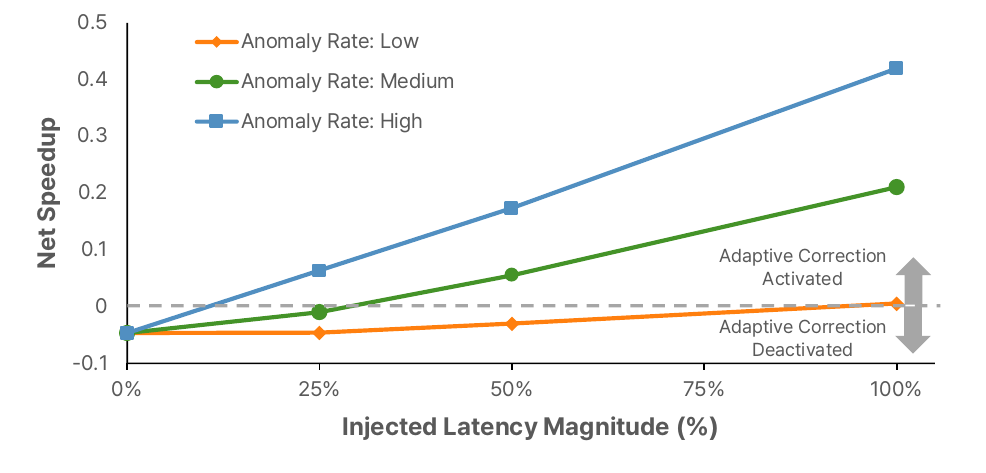}
    \caption{Cost-benefit analysis of Adaptive Correction on LLaVA-OV (Llama-3 8B) using a 4-node cluster with 8xA100 GPUs per node. Net speedup denotes the correction gain minus profiling overhead, evaluated across varying anomaly rates (low, medium, high) representing the outlier frequency and injected latencies relative to the maximum stage duration. The mechanism deactivates when the net speedup is negative to prevent performance degradation.}
    % \vspace{-5pt}
    % \kp{make the graph smaller with smaller texts}
    \label{fig:run_adap}
\end{figure}

% \begin{figure}[t]
%     \centering
%     \includegraphics[width=1\columnwidth,trim={15pt 12pt 15pt 8pt}, clip]{figures/run_adap_3_.pdf}
%     \caption{Cost-benefit analysis of Adaptive Correction on LLaVA-OV (Llama-3 8B) using a 4-node cluster with 8xA100 GPUs per node. Net speedup denotes the correction gain minus profiling overhead, evaluated across varying anomaly rates (low, medium, high) representing the outlier frequency and injected latencies relative to the maximum stage duration. The mechanism deactivates when the net speedup is negative to prevent performance degradation.}
%     % \kp{make the graph smaller with smaller texts}
%     \label{fig:run_adap}
% \end{figure}

To evaluate how effectively each system utilizes the computational capacity of the GPUs assigned to different modules, this experiment analyzes the distribution of throughput for each pipeline stage. We calculate the stage-wise throughput by dividing the total computational load of a stage by its pure computation time, which is defined as the total iteration time minus any idle time attributable to pipeline bubbles. The resulting distributions, visualized in \autoref{fig:stage_thr}, show that the average stage throughput in \sysname is substantially higher than in the baseline systems, while the variance of throughput across its stages is significantly lower. This indicates that \sysname not only achieves a higher level of performance for each stage but also maintains a consistent and balanced throughput across the entire pipeline. This outcome is a direct result of our optimizer, which explicitly selects a 3D parallelism strategy to both maximize the expected throughput for each stage and balance performance across the architecturally distinct modality encoder and LLM modules. In contrast, the baseline systems employ statically configured parallelism that does not account for input-dependent throughput variations, resulting in both a lower average throughput for each stage and a significant performance imbalance across the pipeline.
\begin{table}[t!]
\centering
\caption{Total training time and \sysname overhead across different MLLMs on the mixed dataset using an 8-node cluster with 8×A100 GPUs per node. The relative overhead is the ratio of system overhead to the total training time.}
\label{tab:training_overhead}
\setlength{\tabcolsep}{4pt}
\scalebox{0.75}{
\renewcommand{\arraystretch}{1.5}
\begin{tabular}{lcccc}
    \hline
    \makecell{\textbf{Models}} &
    \makecell{\textbf{Total Training} \\ \textbf{Time (hours)}} & 
    \makecell{\textbf{\sysname} \\ \textbf{Overhead (min)}} & 
    \makecell{\textbf{Relative} \\ \textbf{Overhead (\%)}} \\
    \hline
    LLaVA-OV (Qwen-2.5 7B)  & 6.12   & 7.58 & 2.1 \\
    LLaVA-OV (Llama-3 8B)   & 6.82   & 7.55 & 1.8 \\
    LLaVA-OV (Qwen-2.5 32B) & 11.62   & 9.09 & 1.3 \\
    LLaVA-OV (Llama-3 70B)  & 28.94  & 10.04 & 0.5 \\
    LLaVA-OV (Qwen-2.5 72B) & 30.62  & 9.87 & 0.5 \\
    InternVL (Qwen-2.5 72B) & 34.55 & 9.88 & 0.5 \\
    \hline
\end{tabular}
}
% \vspace{-15pt}
\end{table}

\subsubsection{Cost-Benefit of Adaptive Correction} 
\label{sec:adap_exp}
\revisionnote{R1.O2}\revision{}{This experiment characterizes the activation and deactivation behavior of the Adaptive Correction mechanism across various workload scenarios. The objective is to identify the specific conditions—defined by the frequency of edge cases and the magnitude of their deviation—under which the system toggles the monitoring mechanism. The evaluation employs a cost-benefit analysis with synthetic delays injected into a subset of input shapes, where workload anomalies are categorized by rate—low (1\%), medium (3\%), and high (5\%)—and injected latencies are quantified as a percentage of the observed maximum stage duration.

The evaluation metric, net speedup, is defined as the correction gain minus the profiling overhead ($\sim$4\%). As illustrated in \mbox{\autoref{fig:run_adap}}, the net speedup scales positively with both the anomaly rate and latency magnitude. The results clearly show the operational boundaries of the mechanism; for instance, in scenarios where the benefit is insufficient—such as a low anomaly rate with latency magnitudes $\le$ 50\% or a medium rate with $\le$ 25\%—the mechanism remains deactivated. This confirms that the system successfully manages the monitoring toggle, ensuring that Adaptive Correction is engaged only when the predicted performance gain justifies the overhead.}

\subsubsection{\sysname Overhead}
\label{sec:overhead}

The computational overhead of \sysname is evaluated through two distinct analyses. The first subsection quantifies the one-time initialization cost, encompassing both the offline profiling phase and the strategy generation by the Data-aware 3D Parallelism Optimizer. The subsequent subsection investigates the execution overhead of the Data-aware 3D Parallelism Optimizer and the Online Microbatch Scheduler with respect to resource scaling, assessing their efficiency under large-scale configurations.

\mypar{One-time Profiling Overhead}
The aggregate initialization overhead of \sysname ranges from 7 to 10 minutes across configurations, primarily driven by the Profiling Engine, as detailed in \autoref{tab:training_overhead}. \revisionnote{R1.Q1}\revision{}{Experimental results indicate individual execution times of 6 to 10 minutes for throughput profiling, 3 to 9 minutes for memory profiling, and 1.45 to 1.62 minutes for the Data Profiler. Since these components execute concurrently, the effective profiling overhead is determined by the maximum duration among them, corresponding to the throughput profiling phase.} Finally, the Data-aware 3D Parallelism Optimizer generates the strategy in approximately 3.7 ms. Consequently, the total \sysname overhead constitutes at most 2.1\% of the end-to-end training duration, demonstrating that significant performance gains are achieved with minimal amortized cost.

\begin{figure}[t]
    \centering
    % --- (a) Left Subfigure: Optimizer Table ---
    \begin{subfigure}[t]{0.48\columnwidth}
        \centering
        \includegraphics[width=\columnwidth, trim={0pt 5pt 10pt 5pt}, clip]{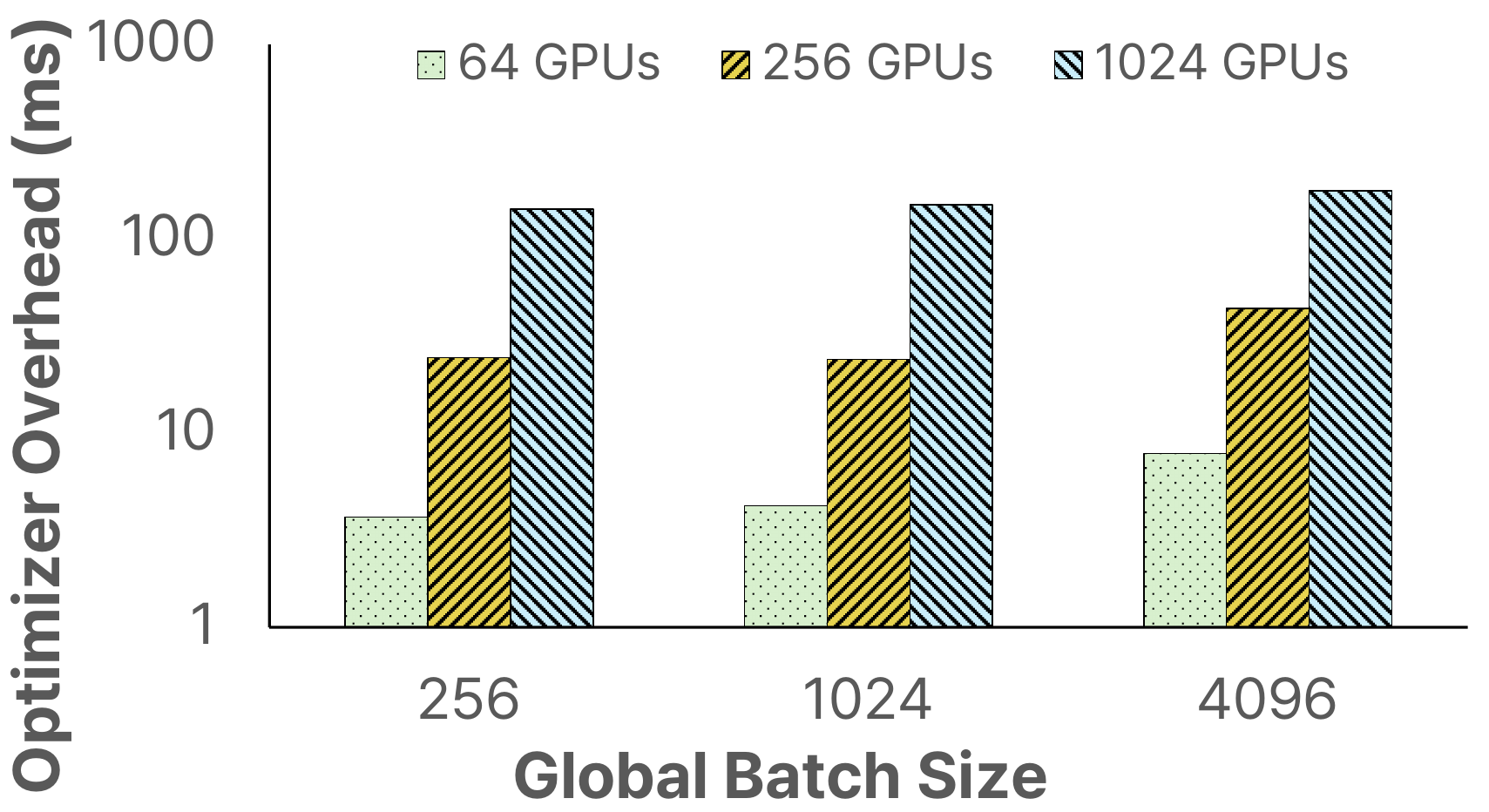}
        \caption{Overhead of the Data-aware 3D Parallelism Optimizer across varying global batch sizes and numbers of GPUs.}
        \label{fig:data_opt_scale}
    \end{subfigure}
    \hfill
    % --- (b) Right Subfigure: Scheduler Graph ---
    \begin{subfigure}[t]{0.48\columnwidth}
        \centering
        \includegraphics[width=\columnwidth, trim={0pt 5pt 10pt 5pt}, clip]{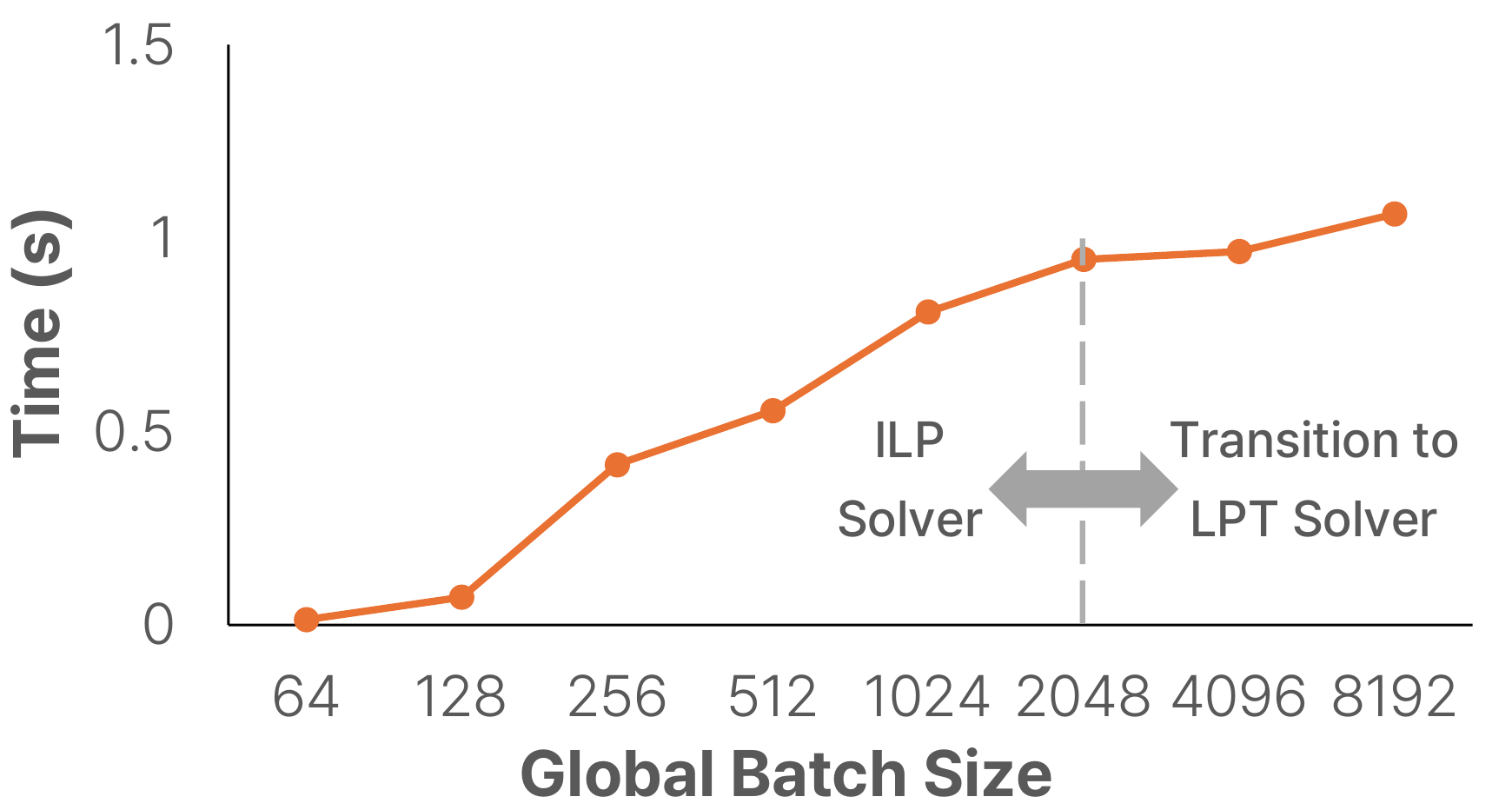}
        \caption{Overhead of the Online Microbatch Scheduler across varying global batch sizes.}
        \label{fig:sched_overhead_graph}
    \end{subfigure}
    
    % \vspace{-5pt}
    \caption{Overhead analysis of \sysname components. The Data-aware 3D Parallelism Optimizer maintains negligible overhead ($<200$ ms) even at large scales (1024 GPUs). The Online Microbatch Scheduler overhead is fully overlapped with computation via asynchronous prefetching, given that training iterations span tens to hundreds of seconds.}
    % \kp{update graph titles}
    \label{fig:system_overhead}
    \vspace{-15pt}
\end{figure}

\mypar{Overhead Sensitivity to Scale} \revisionnoteCommon{R1.O3}\revisionCommon{}{The impact of resource scaling }\revisionnoteCommon{R2.O2}\revisionCommon{}{on the execution latency of the Data-aware 3D Parallelism Optimizer}\revisionnoteCommon{R2.O5}\revisionCommon{}{ and Online Microbatch Scheduler is investigated under } \revisionnoteCommon{R2.O6}\revisionCommon{}{large-scale configurations. As illustrated in \mbox{\autoref{fig:data_opt_scale}}, the execution time for }\revisionnoteCommon{R2.O7}\revisionCommon{}{the Data-aware 3D Parallelism Optimizer remains largely}\revisionnoteCommon{R3.O3}\revisionCommon{}{ invariant to the number of GPUs and global batch size, staying within hundreds of milliseconds. Given that subsequent training runs typically span hours to days, the amortized cost of this preceding offline process is negligible.

As depicted in \mbox{\autoref{fig:sched_overhead_graph}}, the execution time for the Online Microbatch Scheduler exhibits a dependency on global batch size. At a global batch size of 2048, the ILP solver reaches its preset time limit and triggers a fallback to the LPT solver. Despite this transition, the resulting load imbalance deviates by less than 1\% from the theoretical lower bound. Crucially, the asynchronous design ensures that this scheduling latency does not impact the critical path. Since the scheduler operates on the CPU to prefetch data for the subsequent step while a single training iteration consumes tens to hundreds of seconds, the scheduling latency of approximately one second is fully overlapped with model execution.}

\section{Related Work}
\label{sec:related_work}
\revisionnote{R3.O4}\revision{}{Prior research has explored automated and hybrid parallelization strategies, yet they often fall short in the context of MLLMs. Several approaches focus on general matrix-based algorithms or data preprocessing~\cite{hybrid_strategies, uplift, optimizingdatapipeline, tqp, raven}, rather than the specific pipeline constraints of deep learning. Similarly, automating parallelism based on static computational graphs~\cite{alpa, galvatron} assuming uniform batches lacks the runtime adaptivity required for MLLMs, where input-dependent throughput causes severe imbalance.}

Widely adopted distributed learning systems such as DeepSpeed \cite{DeepSpeed}, PyTorch~\cite{pytorch}, and Megatron-LM~\cite{megatron-lm} provide robust 3D parallelism implementations (e.g., ZeRO, kernel fusion) but remain fundamentally data-agnostic. These frameworks apply a monolithic strategy that fails to account for the diverse data characteristics and heterogeneous module architectures inherent to MLLM training, often leading to suboptimal performance in multimodal settings.
% Training large-scale models relies on distributed learning systems to manage computational and memory demands. Widely adopted open-source frameworks include DeepSpeed \cite{DeepSpeed}, PyTorch \cite{pytorch}, and Megatron-LM \cite{megatron-lm}. DeepSpeed utilizes the Zero Redundancy Optimizer (ZeRO) to partition model states across GPUs, reducing memory usage for parameters but not activations, often necessitating throughput-reducing techniques like offloading \cite{zero_offload} for large activation footprints. PyTorch provides foundational primitives for 3D parallelism, combining data, tensor, and pipeline parallelism. Building on PyTorch, Megatron-LM offers highly optimized 3D parallelism implementations, incorporating techniques like kernel fusion and overlapping communication with computation. While effective for single-modality models, these 3D parallelism systems often suffer performance degradation with MLLMs because they are fundamentally data-agnostic and do not account for the diverse data characteristics or heterogeneous module architectures inherent to MLLMs. 
Recent research has explored systems specifically for large-scale multimodal models. DistMM \cite{distmm} determines 3D parallelism based on the computational ratio between modality and language models, but its approach is limited to parallel architectures like CLIP \cite{clip}, unlike the sequential modality-encoder-to-LLM structure common in modern MLLMs. Other works like Optimus \cite{optimus} dynamically schedule encoders into idle stages, and DistTrain \cite{disttrain} uses disaggregated execution with input reordering. Although these studies address MLLM parallelism, they generally do not comprehensively tackle the intertwined challenges of non-uniform computation time and input-dependent throughput variability arising from MLLM heterogeneity. Furthermore, these referenced systems are closed-source, limiting their extensibility and impact on broader research development. In contrast, \sysname addresses these dual challenges through its integrated, data-driven optimization approach, and its open-source nature facilitates future research and development in the community.

\section{Conclusion}
In this paper, we present \sysname, a data-driven framework that optimizes the MLLM training pipeline. \sysname addresses the fundamental inefficiency of data-blind distributed training by explicitly modeling the interaction between input data characteristics and parallel execution behavior. Through its three integrated components--the Profiling Engine, Data-aware 3D Parallelism Optimizer, and Online Microbatch Scheduler--\sysname jointly optimizes parallelism configuration and runtime scheduling to mitigate stage imbalance and input-dependent variability. Our extensive experiments show that \sysname achieves up to \maxgain {\small$\times$} faster training throughput compared to state-of-the-art frameworks such as PyTorch and Megatron-LM, while significantly improving GPU utilization. In summary, \sysname bridges the gap between data diversity and distributed system efficiency, setting a new direction for data-driven optimization in large-scale multimodal model training.

\begin{acks}
This research was supported by NRF grants (No. RS-2025-16068623, No. RS-2024-NR121334), the Korea Basic Science Institute (National Research Facilities and Equipment Center) grant funded by the Ministry of Science and ICT (No. RS-2024-00403860), Advanced Database System Infrastructure (NFEC-2024-11-300458), and the Yonsei University Research Fund (2025-22-0057). We also thank SK Telecom for their dedicated support in providing cloud infrastructure with A100 GPU clusters, which enabled the large-scale experiments in this work.

%This work was conducted in collaboration between the BDAI Lab at Yonsei University and SK Telecom. One author contributed to this work in a personal capacity; the views expressed are solely those of the authors and do not necessarily reflect those of their employer.

%This work was conducted in collaboration among the BDAI Lab at Yonsei University, Microsoft Gray Systems Lab, and SK Telecom, whose collaboration and technical insights significantly contributed to this research.
\end{acks}

% \section*{Acknowledgments-Test}

% Identification of funding sources and other support, and thanks to
% individuals and groups that assisted in the research and the
% preparation of the work should be included in an acknowledgment
% section, which is placed just before the reference section in your
% document.

% This section has a special environment:
% \begin{verbatim}
%   \begin{acks}
%   ...
%   \end{acks}
% \end{verbatim}
% so that the information contained therein can be more easily collected
% during the article metadata extraction phase, and to ensure
% consistency in the spelling of the section heading.

% Authors should not prepare this section as a numbered or unnumbered {\verb|\section|}; please use the ``{\verb|acks|}'' environment.

\bibliographystyle{ACM-Reference-Format}
\bibliography{reference}

\end{document}